\documentclass[journal=jacsat,manuscript=article,layout=twocolumn]{achemso}

\usepackage{chemformula} 
\usepackage[T1]{fontenc} 
\usepackage{multimedia}
\usepackage{tikz}
\usepackage[T1]{fontenc}
\usepackage{epstopdf}
\usepackage{pgfplots}
\usepackage[font={small}]{caption}
\usepackage{ifthen}
\usepackage[singlelinecheck=off,font=bf]{subcaption}
\usepackage{tikz,siunitx,mwe}
\usepackage{pgfplots}
\usepackage{pgfplotstable}
\usepackage{upgreek}
\usepackage{amsmath}
\usepackage{mathtools}
\usepackage{circuitikz}
\usepackage{enumerate}
\usepackage{enumitem}

\usetikzlibrary{decorations.markings,calc}
\newcount\myangle  

\usetikzlibrary{calc}

\tikzstyle{block} = [draw, fill=blue!20, rectangle, 
minimum height=3em, minimum width=6em]
\tikzstyle{sum} = [draw, fill=blue!20, circle, node distance=1cm]
\tikzstyle{input} = [coordinate]
\tikzstyle{output} = [coordinate]
\tikzstyle{pinstyle} = [pin edge={to-,thin,black}]

\newcommand{\executeiffilenewer}[3]{%
	\ifnum\pdfstrcmp{\pdffilemoddate{#1}}%
	{\pdffilemoddate{#2}}>0%
	{\immediate\write18{#3}}\fi%
}

\graphicspath{{figs/}}

\pgfplotsset{compat=1.6}
\definecolor{darkred}{RGB}{0.5,0,0}
\usetikzlibrary{fadings}

\let\oldmaketitle\maketitle
\let\maketitle\relax

\usetikzlibrary{hobby}   
\definecolor{aqua}{rgb}{0.0, 1.0, 1.0}
\definecolor{ao(english)}{rgb}{0.0, 0.5, 0.0}

\author{Mathieu Giroux}
\affiliation[University of Ottawa]
{Department of Mechanical Engineering, University of Ottawa, Ottawa, Ontario, Canada}
\author{Chang Zhang}
\affiliation[University of Ottawa]
{Department of Mechanical Engineering, University of Ottawa, Ottawa, Ontario, Canada}
\author{Nikaya Snell}
\affiliation[University of Ottawa]
{Department of Mechanical Engineering, University of Ottawa, Ottawa, Ontario, Canada}
\author{Gengyang Mu}
\affiliation[University of Ottawa]
{Department of Mechanical Engineering, University of Ottawa, Ottawa, Ontario, Canada}
\author{Michel Stephan}
\affiliation[University of Ottawa]
{Department of Mechanical Engineering, University of Ottawa, Ottawa, Ontario, Canada}
\author{Raphael St-Gelais}
\affiliation[University of Ottawa]
{Department of Mechanical Engineering, University of Ottawa, Ottawa, Ontario, Canada}
\email{raphael.stgelais@uottawa.ca}
\title[Near-Field Heat Transfer Measurement]
  {High Resolution Measurement of\\ Near-Field Radiative Heat Transfer enabled\\ by Nanomechanical Resonators}

\abbreviations{IR,NMR,UV}
\keywords{American Chemical Society, \LaTeX}

\begin{document}

\fontsize{11}{12}\selectfont
\twocolumn[
\begin{@twocolumnfalse}
	\oldmaketitle
	\begin{abstract}
		Near-field radiative heat transfer (NFHT) research currently suffers from an imbalance between numerous theoretical studies, as opposed to experimental reports that remain, in proportion, relatively scarce. Existing experimental platforms all rely on unique custom-built devices on which it is difficult to integrate new materials and structures for studying the breadth of theoretically proposed phenomena. Here we show high-resolution NFHT measurements using, as our sensing element, silicon nitride (SiN) freestanding nanomembranes—a widely available platform routinely used in materials and cavity optomechanics research. We measure NFHT by tracking the high mechanical quality (Q) factor ($>2\times10^6$) resonance of a membrane placed in the near-field of a hemispherical hot object. We find that high Q-factor enables a temperature resolution ($1.2\times10^{-6} \ \mathrm{K}$) that is unparalleled in previous NFHT experiments. Results are in good agreement with a custom-built model combining heat transport in nanomembranes and the effect of non-uniform stress/temperature on the resonator eigenmodes. 
	\end{abstract}
\end{@twocolumnfalse}
]
Near-field radiative heat transfer (NFHT) has attracted a lot of attention over the past years for potential applications such as energy conversion \cite{Laroche_2006,Milovich_2020,Fiorino_2018,DiMatteo_2003,Bhatt_2020}, active thermal control \cite{Zhu_2013,Biehs_2011,Ben-Abdallah_2015,Guha_2012}, or for more fundamental studies on the effect of various material platforms \cite{Biehs_2007,Salihoglu_2019,Basu_2015,Kim_2015,Francoeur_2008}, material geometries \cite{Jin_2019}, or external fields \cite{Moncada-Villa_2015}. Near-field thermal radiation consists of evanescent electromagnetic coupling occurring between two bodies at subwavelength distances, greatly increasing the radiative exchange beyond conventional laws of thermal radiation \cite{Tien_2002,Cravalho_1967,Polder_1971}. As demonstrated theoretically \cite{Molesky_2015,Ilic_2012}, this enhancement could allow conversion of heat to electricity with a higher efficiency than any existing portable solid-state solution. In addition, by concentrating heat into a narrow spectral bandwidth \cite{Pendry_1999,Carminati_1999,Volokitin_2001,Babuty_2013}, near-field thermal radiation could be used to better control heat flux \cite{Jin_2019}, thus potentially enabling active thermal control devices such as thermal transistors \cite{Ben-Abdallah_2015,Ben-Abdallah_2014}.

While lots of theoretical work has been reported on near-field thermal radiation, experimental progress is relatively scarce due to important technical challenges, i.e., maintaining objects at small distances while avoiding contact, under important temperature gradients. Different approaches have been reported over the years. Some rely on nanotips \cite{Guha_2012,Kim_2015,Kloppstech_2017} or microspheres \cite{Lucchesi_2019,Van_Zwol_2012,Song_2015,Shen_2012,Rousseau_2009,Menges_2016}  to avoid alignment issues, others rely on more complex systems using active parallelism control \cite{Ottens_2011,Shi_2019,Lim_2018,Kralik_2012,Ijiro_2015}, and others use static systems with spacers \cite{Sabbaghi_2020,Tang_2020,DiMatteo_2001,Hu_2008,Ito_2015}. Current experimental platforms all rely on unique, custom-built systems, on which it is difficult to integrate new materials to experimentally study and confirm existing theoretical work on NFHT. There is consequently a substantial imbalance between a large body of theoretically investigated phenomena \cite{Laroche_2006,Milovich_2020,Zhu_2013,Biehs_2011,Ben-Abdallah_2015,Biehs_2007,Salihoglu_2019,Basu_2015,Francoeur_2008,Jin_2019,Cravalho_1967,Polder_1971,Molesky_2015,Ilic_2012,Pendry_1999,Carminati_1999,Volokitin_2001,Babuty_2013,Zhao_2017}, and of relatively modest experimental capabilities. Due do their highly custom nature, experimental platforms are typically reported only once in scientific literature and are rarely used again as a systematic tool for studying other NFHT effects. Moreover, a large proportion of these platforms rely on \ch{SiO2} \cite{Lucchesi_2021}, meaning that a lot of materials with attractive properties such as graphene \cite{Nika_2012}, metals \cite{Biehs_2007}, multilayers \cite{Shi_2017}, lossy materials \cite{Jin_2019}, hyperbolic materials \cite{Salihoglu_2019,Moncada-Villa_2015}, and metamaterials \cite{Basu_2015} remain largely untested.

\begin{figure}[!htb]
		\begin{subfigure}{\columnwidth}
			\centering
			\begin{tikzpicture}
			\fill [blue,opacity=1] (-1.2,2.2)--(-1.2,2.15)--(1.1,2.15)--(1.1,2.2)--cycle;
			\draw [line width=0.08] (-1.2,2.2)--(-1.2,2.15)--(1.1,2.15)--(1.1,2.2)--cycle;
			\draw [line width=0.08] (0.15,2.325) arc(73.1:106.9:0.7);
			\draw [line width=0.08] (0.05,2.275) arc(73.1:106.9:0.36);
			\draw [line width=0.08] (-0.25,2.025) arc(253.1:286.9:0.7);
			\draw [line width=0.08] (-0.15,2.075) arc(253.1:286.9:0.36);
			\fill[gray, opacity=0.3] (-1.2,2.15)--(-0.65,2.15)--(-0.65,2.005)--(-0.85,1.62)--(-1.2,1.62)--(-1.2,2.15);
			\draw[] (-1.2,2.15)--(-0.65,2.15);
			\draw[](-0.65,2.15)--(-0.65,2.005);
			\draw[](-0.65,2.005)--(-0.85,1.62);
			\draw[](-0.85,1.62)--(-1.2,1.62);
			\draw[](-1.2,1.62)--(-1.2,2.15);
			\fill[gray, opacity=0.3] (1.1,2.15)--(0.55,2.15)--(0.55,2.005)--(0.75,1.62)--(1.1,1.62)--(1.1,2.15);
			\draw[] (1.1,2.15)--(0.55,2.15);
			\draw[](0.55,2.15)--(0.55,2.005);
			\draw[](0.55,2.005)--(0.75,1.62);
			\draw[](0.75,1.62)--(1.1,1.62);
			\draw[](1.1,1.62)--(1.1,2.15);
			\draw (1,1.62)--(0.85,1.62)--(0.85,1.47)--(1,1.47)--(1,1.62);
			\fill[black, opacity=1] (1,1.62)--(0.85,1.62)--(0.85,1.47)--(1,1.47)--(1,1.62);
			\draw[line width=0.08] (0.85,1.42) arc(265:185:0.08);
			\draw[line width=0.08] (0.84,1.39) arc(265:185:0.1);
			\draw[line width=0.08] (1,1.42) arc(275:355:0.08);
			\draw[line width=0.08] (1.01,1.39) arc(275:355:0.1);
			\draw[dash pattern=on 6pt off 1.8pt ](-5,-0.35)--(-5,-0.65)--(0.925,-0.65)--(0.925,1.47);
			\fill [black,opacity=1] (-3.4,0.75)--(-0.07,0.75)--(-0.07,0.79)--(-3.4,0.79)--cycle;
			\fill [black,opacity=1] (-2.9,0.77)--(-3.0,0.87)--(-3.0,0.67)--cycle;
			\fill [black,opacity=1] (-0.25,0.77)--(-0.35,0.87)--(-0.35,0.67)--cycle;
			\fill [black,opacity=1] (-0.85,0.77)--(-0.75,0.87)--(-0.75,0.67)--cycle;
			\fill [black,opacity=1] (-0.07,0.75)--(-0.03,0.75)--(-0.03,1.9)--(-0.07,1.9)--cycle;
			\draw [line width=0.08] (-3.4,0.72)--(-3.4,0.82)
			(-3.4,0.82)--(-3.5,0.82)
			(-3.4,0.72)--(-3.5,0.72)
			(-3.5,0.82)--(-3.5,0.97)
			(-3.5,0.72)--(-3.5,0.57)
			(-3.5,0.97)--(-4.4,0.97)
			(-3.5,0.57)--(-4.4,0.57)
			(-4.4,0.97)--(-4.4,0.57);
			\fill [blue,opacity=0.25] 
			(-3.4,0.72)--(-3.4,0.82)--(-3.5,0.82)--(-3.5,0.97)--(-4.4,0.97)--(-4.4,0.57)--(-3.5,0.57)--(-3.5,0.72)--(-3.4,0.72);
			\node[font=\fontsize{10pt}{12}\selectfont] at (-3.95,0.77) {Laser};
			\draw[line width=0.08] (-2.31,0.87)--(-1.31,0.87)
			(-2.31,0.87)--(-2.31,0.6)
			(-2.31,0.6)--(-1.31,0.6)
			(-1.31,0.6)--(-1.31,0.87);
			\fill [blue,opacity=0.1] 
			(-2.31,0.87)--(-2.31,0.6)--(-1.31,0.6)--(-1.31,0.87)--(-2.31,0.87);
			\node[font=\fontsize{10pt}{12}\selectfont] at (-1.81,1.03) {90:10 coupler};
			\fill [black,opacity=1] (-3.22,0.15) arc(135:45:2)--(-0.4,0.15)--(-0.364644661,0.185355339) arc(45:136:2.05)--cycle;
			\fill [black,opacity=1] (-0.38,0.17) circle (0.05);
			\node[font=\fontsize{10pt}{12}\selectfont] at (-0.4,-0.1) {Unused};
			\fill [black,opacity=1] (-2.8,0.52)--(-2.75,0.62)--(-2.7,0.47)--cycle;
			\fill [black,opacity=1] (-0.75,0.46)--(-0.78,0.58)--(-0.87,0.43)--cycle;
			\draw[line width=0.08](-1.5,0.15)--(-3.8,0.15);
			\draw[line width=0.08](-1.5,0.15)--(-1.5,-0.35);
			\draw[line width=0.08](-1.5,-0.35)--(-3.8,-0.35);
			\draw[line width=0.08](-3.8,-0.35)--(-3.8,0.15);
			\node[font=\fontsize{10pt}{12}\selectfont] at (-2.65,-0.1) {Photodetector};
			\fill [blue,opacity=0.25] 
			(-1.5,0.15)--(-1.5,-0.35)--(-3.8,-0.35)--(-3.8,0.15)--(-1.5,0.15);
			\draw[dash pattern=on 6pt off 1.8pt ](-3.8,-0.1)--(-4.2,-0.1);
			\draw[line width=0.08](-4.2,0.35)--(-5.7,0.35);
			\draw[line width=0.08](-4.2,0.35)--(-4.2,-0.35);
			\draw[line width=0.08](-4.2,-0.35)--(-5.7,-0.35);
			\draw[line width=0.08](-5.7,-0.35)--(-5.7,0.35);
			\node[line width=0.08][font=\fontsize{10pt}{12}\selectfont] at (-4.95,0.175) {Lock-in};
			\node[font=\fontsize{10pt}{12}\selectfont] at (-4.95,-0.175) {Amplifier};
			\fill [blue,opacity=0.25] 
			(-4.2,0.35)--(-4.2,-0.35)--(-5.7,-0.35)--(-5.7,0.35)--(-4.2,0.35);
			\draw[line width=0.08] (-0.65,3.2)--(-0.65,3.25)
			(-0.65,3.25)--(0.55,3.25)
			(0.55,3.25)--(0.55,3.2);
			\fill [orange,opacity=1] (0.55,3.2)--(-0.65,3.2)--(-0.65,3.25)--(0.55,3.25)--cycle;
			\shade [inner color=orange, outer color=white, even odd rule](-0.75,3.35)--(-0.75,3.1)--(0.65,3.1)--(0.65,3.35)--cycle (-0.65,3.25)--(-0.65,3.2)--(0.55,3.2)--(0.55,3.25)--cycle;
			\draw[] (-1.1991165,3.1)--(-1.1991165,3.2);
			\draw[] (1.0991165,3.1)--(1.0991165,3.2);
			\draw[] (-1.1991165,3.2)--(1.0991165,3.2);
			\draw[] (1.0991165,3.1)--(-1.1951165,3.1) arc(220:320:1.49);
			\fill[gray, opacity=0.1] (-1.1991165,3.1) arc(220:320:1.5)--(1.0991165,3.2)--(-1.1991165,3.2)--cycle;
			\draw[latex-latex, line width=0.6pt, red]	(-1.35,2.7)--(-1.35,3.3);	
			\draw[-latex, line width=0.6pt]	(-3.7,1.4)--(-1.2,2.175);
			\node[above, font=\fontsize{10pt}{12}\selectfont] at (-4.5,1.1) {SiN Layer};
			\draw[-latex, line width=0.6pt]	(2.05,1.5)--(1.1,1.87);
			\node[below, font=\fontsize{10pt}{12}\selectfont] at (2.2,1.75) {Si}; 
			\node[below, font=\fontsize{10pt}{12}\selectfont] at (2.2,1.45) {Frame};
			\draw[-latex, line width=0.6pt]	(1.85,2.8)--(1.0,3);
			\node[above, font=\fontsize{10pt}{12}\selectfont] at (2.175,2.5) {Half};
			\node[above, font=\fontsize{10pt}{12}\selectfont] at (2.175,2.15) {Sphere};
			\draw[-latex, line width=0.6pt]	(0.25,3.8)--(0,3.25);
			\node[above, font=\fontsize{10pt}{12}\selectfont] at (0.25,3.7) {Heater};
			\draw[-latex, line width=0.6pt]	(-1,3.8)--(-0.875,3.25);
			\node[above, font=\fontsize{10pt}{12}\selectfont] at (-1,3.7) {RTD};
			\draw[-latex, line width=0.6pt]	(1.8,0.65)--(1.1,1.4);
			\node[below, font=\fontsize{10pt}{12}\selectfont] at (2.2,0.75) {Piezo};
			\draw[-latex, line width=0.6pt]	(1.6,0.1)--(-0.03,1.4);
			\node[below, font=\fontsize{10pt}{12}\selectfont] at (2.12,0.2) {Optical};
			\node[below, font=\fontsize{10pt}{12}\selectfont] at (2.12,-0.15) {Fiber};
			\draw[dashed,line width=0.75,dash pattern=on 2.0pt off 0.6pt](1.4,3.4)--(-1.5,3.4)--(-1.5,1.35)--(1.4,1.35)--(1.4,3.4);
			\draw[dashed,line width=0.75,dash pattern=on 2.0pt off 0.6pt](-5.45,3.85)--(-5.7,3.85);
			\node[font=\fontsize{10pt}{12}\selectfont] at (-4.75,3.85) {Vacuum};
			\draw[-latex, line width=0.6pt]	(-3.55,2.4)--(-1.35,3);
			\node[font=\fontsize{10pt}{12}\selectfont] at (-4.5,2.4) {Half Sphere};
			\node[font=\fontsize{10pt}{12}\selectfont] at (-4.5,2.1) {Displacement};
			\fill [green,opacity=1] (-0.85,3.2)--(-1.05,3.2)--(-1.05,3.25)--(-0.85,3.25)--cycle;
			\draw[line width=0.08] (-1.05,3.2)--(-1.05,3.25)--(-0.85,3.25)--(-0.85,3.2)--(-1.05,3.2);
			\draw (0.45,3.25) sin (0.9,3.55) cos (1.45,3.7);
			\draw (0.25,3.25) sin (0.85,3.65) cos (1.45,3.85);
			\draw (-1.0,3.25) sin (-1.6,3.55) cos (-2,3.7);
			\draw (-0.9,3.25) sin (-1.45,3.65) cos (-2,3.85);
			\draw[line width=0.08](1.45,4.1)--(2.65,4.1);
			\draw[line width=0.08](2.65,3.25)--(2.65,4.1);
			\draw[line width=0.08](2.65,3.25)--(1.45,3.25);
			\draw[line width=0.08](1.45,3.25)--(1.45,4.1);
			\node[font=\fontsize{10pt}{12}\selectfont] at (2.05,3.85) {Power};
			\node[font=\fontsize{10pt}{12}\selectfont] at (2.05,3.5) {Supply};
			\fill [blue,opacity=0.25] 
			(1.45,4.1)--(1.45,3.25)--(2.65,3.25)--(2.65,4.1)--(1.45,4.1);
			\draw[line width=0.08](-3.95,4.1)--(-3.95,2.9);
			\draw[line width=0.08](-3.95,2.9)--(-2,2.9);
			\draw[line width=0.08](-2,2.9)--(-2,4.1);
			\draw[line width=0.08](-2,4.1)--(-3.95,4.1);
			\node[font=\fontsize{10pt}{12}\selectfont] at (-2.975,3.85) {Wheatstone};
			\node[font=\fontsize{10pt}{12}\selectfont] at (-2.975,3.5) {Bridge};
			\node[font=\fontsize{10pt}{12}\selectfont] at (-2.975,3.15) {Circuit};
			\fill [blue,opacity=0.25] 
			(-3.95,4.1)--(-2,4.1)--(-2,2.9)--(-3.95,2.9)--(-3.95,4.1);
			
			\end{tikzpicture}
			\vspace{-7.5mm}
			\caption{}
			\label{fig_1:sub-first}
		\end{subfigure}
		\begin{subfigure}{\columnwidth}
			
			\begin{tikzpicture}
			\node[anchor=south west,inner sep=0] (image) at (0,0) {\includegraphics[width=\linewidth]{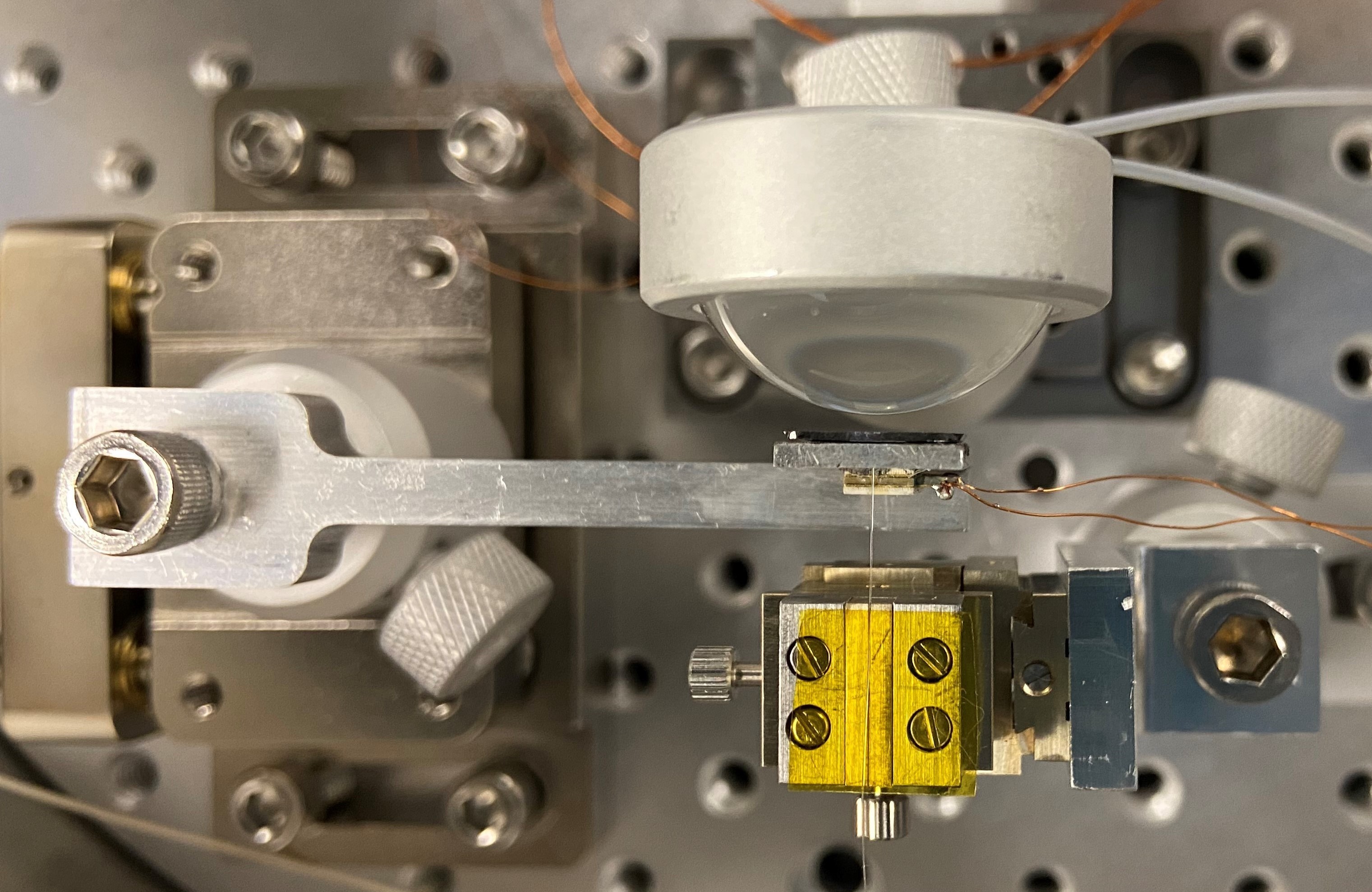}};
			\node[anchor=south west,inner sep=0] (image) at (0,3.8) {\includegraphics[width=0.2\linewidth]{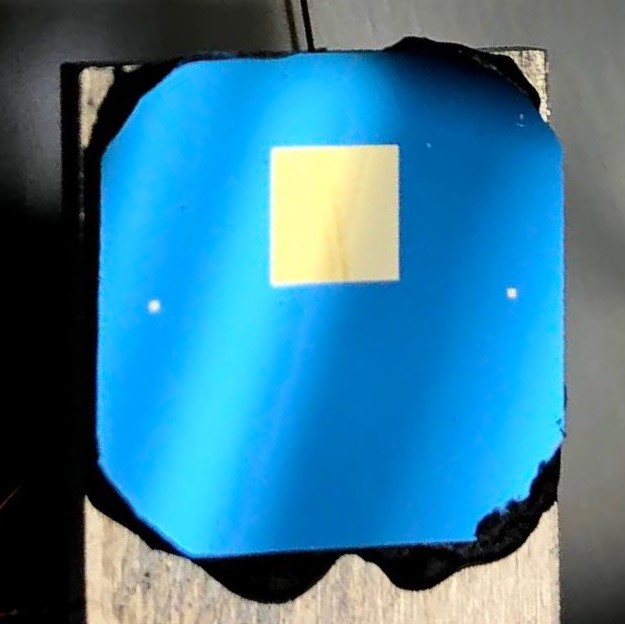}};
			\draw[ultra thick,red] (0.006,3.8)--(1.7,3.8)--(1.7,5.478)--(0.035,5.478)--(0.035,3.8); \draw[thick,dashed,-stealth,red] (1.7,3.8)--(4.85,2.85); 
			\draw[thick,dashed,-stealth,white] (0.6,0.5)--(5.2,2.47); 
			\node[white] at (0.6,0.25) {Piezo};
			\draw[thick,dashed,-stealth,white] (7.125,0.5)--(5.39,2.125);
			\node[white] at (7.125,0.25) {Optical Fiber};
			\draw[thick,dashed,-stealth,white] (2.1,5.3)--(1.1,5.1);
			\node[white] at (3.5,5.3) {SiN Membrane};
			\draw[thick,dashed,-stealth,white] (2.15,4.2)--(1.5,4.5);
			\node[white] at (3,4.2) {Si Frame};
			
			\end{tikzpicture} 
			\vspace{-7.5mm} 
			\caption{}
			\label{fig_1:sub-second}
		\end{subfigure}
		\vspace{-3.5mm}
		\caption{ 
			(\subref{fig_1:sub-first}) Schematic of the experimental NFHT measurement platform.
			(\subref{fig_1:sub-second}) Photograph of the NFHT measurement platform inside the vacuum chamber.}
		\label{fig_1}
	\end{figure}
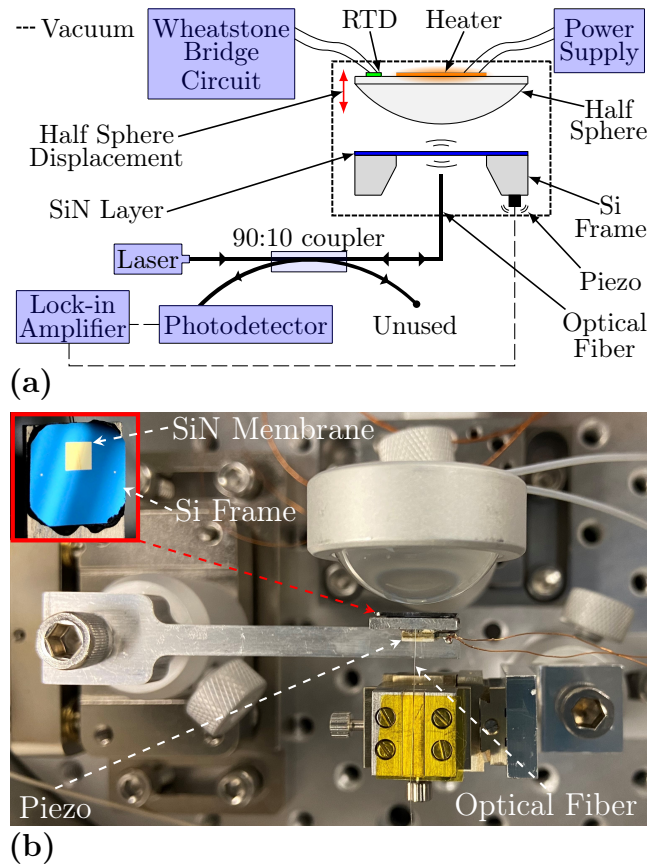

Here we show high-resolution NFHT measurements using, as our sensing element, silicon nitride (SiN) membranes (Fig.~\ref{fig_1})—a widely available platform routinely used in nanomaterials \cite{Creemer_2010} and optomechanics research \cite{Thompson_2008,Wilson_2009,Norte_2016}. Such membranes are commercially available from multiple suppliers, therefore resulting in a reproducible and flexible experimental platform. Moreover, emerging materials are routinely deposited on SiN membranes for transmission electron microscopy (TEM) analysis. Protocols for depositing materials on SiN are therefore well developed, making SiN an ideal substrate for testing new materials for NFHT research.

Our experimental approach, illustrated in Fig.~\ref{fig_1:sub-first}, uses as a hot surface a glass half sphere (Thorlabs BK7 lens, $12.7 \ \mathrm{mm}$ radius) onto which a metal-ceramic heater is attached. The half sphere is used to simplify alignment, allowing the use of a single-axis positioner. The half sphere is brought closer to the SiN membrane using a vacuum positioner actuated with an inertia drive (New Focus Picomotor\textsuperscript{TM}). The temperature of the membrane is measured in real-time with an optical fiber interferometer \cite{Rugar_1989}, by measuring the shift of its mechanical resonance frequency due to thermal expansion \cite{Zhang_2019,Sadeghi_2020}. To suppress fluidic damping and convective heat transfer, the resonator and the glass half sphere are positioned in a custom-designed high vacuum chamber operating at a typical pressure of $2\times10^{-6} \ \mathrm{Torr}$. The membrane used for the current experiment is made of low stress SiN (built-in stress $\mathrm{\sim\!100 \ MPa}$), with nominal dimensions of $3 \ \mathrm{mm}$ in side length and $t_{SiN}=100 \ \mathrm{nm}$ thickness. The membrane was fabricated in-house, but comparable devices can be purchased commercially from various suppliers. The silicon (Si) frame of the membrane is mounted on an aluminum support (see Fig.~\ref{fig_1:sub-second}) using nickel paste (Pelco\textsuperscript{\textregistered}) enabling a good thermal conduction for maintaining the frame at room temperature. A piece of piezoelectric ceramic, mounted on the back of the aluminium support, is used to drive the membrane at its resonance frequency. 

\begin{figure}[!htb]
	\centering
	\begin{subfigure}{\columnwidth}
		\def\svgwidth{\columnwidth}
	\executeiffilenewer{figs/Heating_scan.svg}{figs/Heating_scan.pdf}%
	{inkscape -D figs/Heating_scan.svg  -o figs/Heating_scan.pdf --export-latex}%
	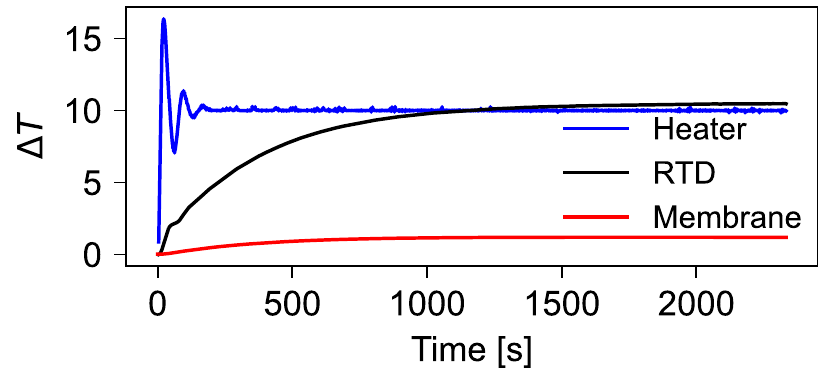
		\vspace{-12mm}
		\caption{}
		\label{fig_2:sub-first}
		\vspace{-0.6mm}
	\end{subfigure}
	\begin{subfigure}{\columnwidth}
		\hspace{-3mm}
		\begin{tikzpicture}
		\node[anchor=south west,inner sep=0] (image) at (0,0) {\def\svgwidth{\columnwidth}
	\executeiffilenewer{figs/picomotor_calibration.svg}{figs/picomotor_calibration.pdf}%
	{inkscape -D figs/picomotor_calibration.svg  -o figs/picomotor_calibration.pdf --export-latex}%
	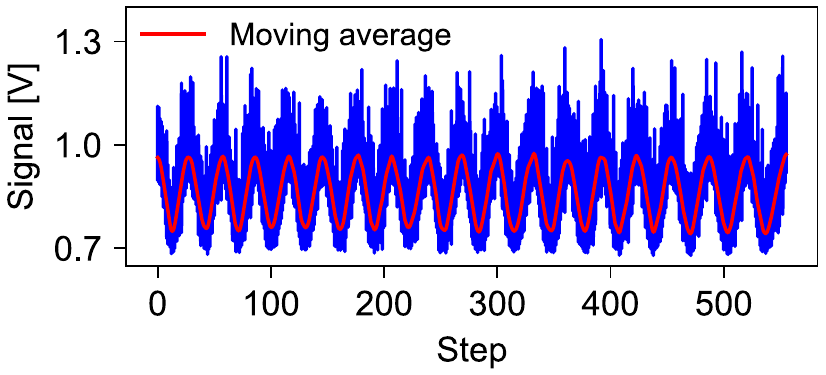
};
		\begin{scope}[shift={(-2.2,-0.98)}]
		\scalebox{1.35}{
			\fill[white] (6.74,2.71)--(6.74,3.75)--(8.5,3.75)--(8.5,2.71)--(6.74,2.71);
			\draw[thick](6.74,3.78)--(6.74,2.71)--(8.52,2.71);
			\clip (6.76,2.73) rectangle + (1.75,1.04);
			\fill [blue,opacity=1,rotate=-90,shift={(-3.2,5.69)}] (-1.2,2.2)--(-1.2,2.15)--(1.1,2.15)--(1.1,2.2)--cycle;
			\draw[line width=0.08,rotate=-90,shift={(-3.2,5.69)}] (-1.2,2.15)--(-0.4,2.15)--(-0.4,2.005)--(-0.6,1.62)--(-1.2,1.62)--(-1.2,2.15);
			\fill[gray, opacity=0.3,rotate=-90,shift={(-3.2,5.69)}] (-1.2,2.15)--(-0.4,2.15)--(-0.4,2.005)--(-0.6,1.62)--(-1.2,1.62)--(-1.2,2.15);
			\draw[line width=0.08,rotate=-90,shift={(-3.2,5.69)}] (1.1,2.15)--(0.3,2.15)--(0.3,2.005)--(0.5,1.62)--(1.1,1.62)--(1.1,2.15);
			\fill[gray, opacity=0.3,rotate=-90,shift={(-3.2,5.69)}] (1.1,2.15)--(0.3,2.15)--(0.3,2.005)--(0.5,1.62)--(1.1,1.62)--(1.1,2.15);
			\fill [aqua,opacity=0.25,shift={(6.51,2.48)}] (0,0.87)--(0,0.67)--(0.8,0.67)--(0.8,0.87)--cycle;
			\fill [yellow,opacity=1,shift={(6.51,2.48)}] (0,0.87)--(0,0.95)--(0.8,0.95)--(0.8,0.87)--cycle;
			\fill [yellow,opacity=1,shift={(6.51,2.48)}] (0,0.67)--(0,0.59)--(0.8,0.59)--(0.8,0.67)--cycle;
			\draw[line width=0.08,shift={(6.51,2.48)}] (0,0.59)--(0.8,0.59)--(0.8,0.95)--(0,0.95);
			\draw[line width=0.08,shift={(6.51,2.48)}] (0,0.67)--(0.8,0.67);
			\draw[line width=0.08,shift={(6.51,2.48)}] (0,0.87)--(0.8,0.87);
			\draw[line width=0.08,rotate=-90,shift={(-3.2,5.69)}] (-1.1991165,3.1)--(-1.1991165,3.2);
			\draw[line width=0.08,rotate=-90,shift={(-3.2,5.69)}] (1.0991165,3.1)--(1.0991165,3.2);
			\draw[line width=0.08,rotate=-90,shift={(-3.2,5.69)}] (-1.1991165,3.2)--(1.0991165,3.2);
			\draw[line width=0.08,rotate=-90,shift={(-3.2,5.69)}] (1.0991165,3.1)--(-1.1991165,3.1) arc(220:320:1.5);
			\fill[gray, opacity=0.1,rotate=-90,shift={(-3.2,5.69)}] (-1.1991165,3.1) arc(220:320:1.5)--(1.0991165,3.2)--(-1.1991165,3.2)--cycle;	
			\draw[{Latex[scale=0.53333]}-,black,shift={(6.51,2.48)}] (0.3,0.81) .. controls (0.925,0.81) and (0.925,0.86) .. (0.4,0.86);
			\draw[{Latex[scale=0.53333]}-,red,shift={(6.51,2.48)}] (0.5,0.76) .. controls (1.598,0.76) and (1.598,0.86) .. (0.4,0.86);
			\draw[{Latex[scale=0.5333]}-,ao(english),shift={(6.51,2.48)}] (0.7,0.71) .. controls (2.113,0.71) and (2.113,0.86) .. (0.4,0.86);
		};
		\end{scope}
		\end{tikzpicture}      
		\vspace{-12mm}
		\caption{}
		\label{fig_2:sub-second}
		\vspace{-3.125mm}
	\end{subfigure}
	
	\caption{(\subref{fig_2:sub-first}) Initial temperature calibration and stabilization before NFHT measurements. The heater temperature (blue) is actively stabilized to $\Delta T= 10 \ \mathrm{K}$. The membrane (in the far-field) and a RTD placed on the half sphere are stable after $t\approx 1000 \ \mathrm{sec}$.  (\subref{fig_2:sub-second}) Interferometric measurement of the half sphere displacement, yielding $25.3 \ \mathrm{nm}$ per actuator step in this particular instance. Inset: schematic of the reflected optical signals at the tip of the optical fiber (black), the resonator surface (red), and the surface of the half sphere (green). }
	\label{fig_2}
\end{figure}
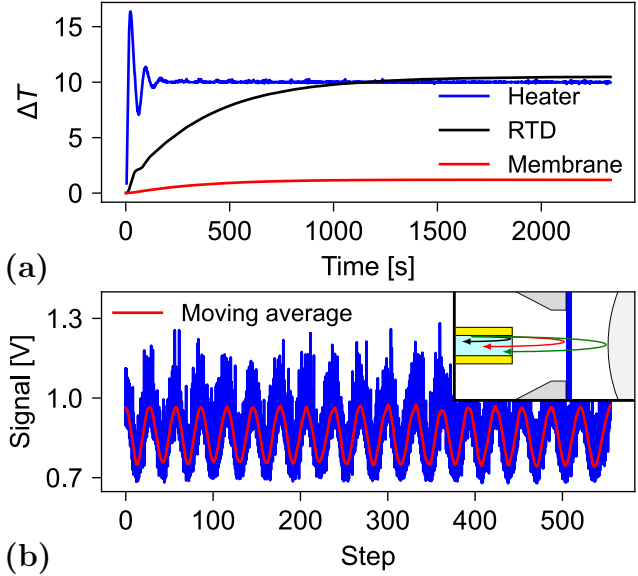

The glass half sphere is heated by a metal-ceramic heater connected to a power supply, which is actively controlled from a computer. We rely on the linear variation of the resistance to track the temperature of the metal-ceramic heater and actively control its temperature with a closed-loop custom PID control algorithm. The temperature of the glass half sphere is also measured with a resistance temperature detector (RTD), which is connected to a modified Wheatstone bridge circuit, relying on the Wien-bridge oscillator principle \cite{Lipoma_1971} to reduce noise. At the beginning of the heating process, the PID adjusts the heater to a stable temperature, which takes approximately 200 seconds (i.e., blue curve in Fig.~\ref{fig_2:sub-first}). After about 1000 seconds, the temperature of the whole system, i.e., the half sphere RTD (black curve) and the membrane (red curve), stabilizes. At this point in time, the heater and RTD temperature are superimposed, thus confirming good thermal contact between the heater, the glass sphere, and the RTD. 

The optical fiber interferometer plays the double role of measuring the membrane resonance frequency \cite{Rugar_1989} (and thus its temperature) and of calibrating the open-loop positioner stage displacement. In a typical implementation of such optical fiber interferometer \cite{Rugar_1989}, reflection of laser signal occurs at two interfaces: the tip of the optical fiber (Fig.~\ref{fig_2:sub-second}, black arrow) and the surface of the resonator (red arrow). However, in our case, reflection also occurs at the surface of the half sphere (green arrow). This creates two interference signals (i.e., fiber-sphere and fiber-resonator) that can easily be demultiplexed, given that the resonator signal is centered at a much higher frequency—typically $83 \ \mathrm{kHz}$ for eigenmode order ($m$, $n$) = (2, 2) with the current resonator dimensions. As the sphere is displaced, the slower fiber-sphere interference signal shows a periodic pattern (Fig.~\ref{fig_2:sub-second}) that allows calibration of the stage displacement, considering our laser (RIO ORION\textsuperscript{TM}) wavelength of $1563.52 \ \mathrm{nm}$. The actuator displacement is measured for every scan, yielding typical values in the $24\! -\!25.5 \ \mathrm{nm/step}$ range. The membrane resonance frequency is measured in real-time by closed-loop phase locking (PLL) of the lock-in amplifier internal oscillator to the membrane resonance mode order ($m$, $n$) = (2, 2) \cite{Zwickl_2008}. 

\begin{figure}[!htb]
	\centering
	\begin{subfigure}{\columnwidth}
		\def\svgwidth{\columnwidth}
	\executeiffilenewer{figs/Raw_data.svg}{figs/Raw_data.pdf}%
	{inkscape -D figs/Raw_data.svg  -o figs/Raw_data.pdf --export-latex}%
	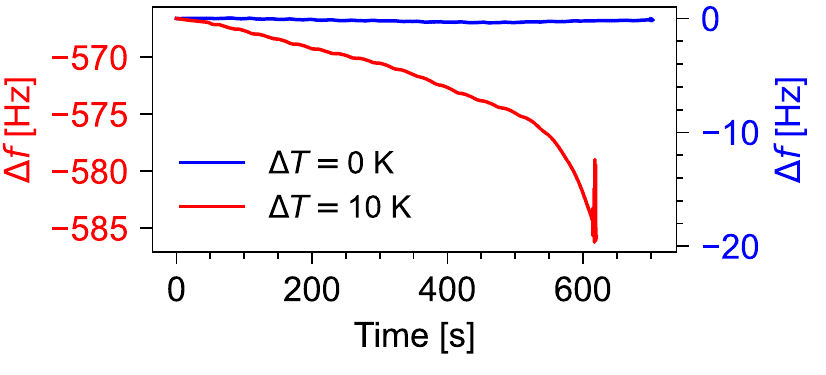
		\vspace{-13.25mm}
		\caption{}
		\label{fig_3:sub-first}
		\vspace{-0.6mm}
	\end{subfigure}
	\begin{subfigure}{\columnwidth}
		\def\svgwidth{\columnwidth}
	\executeiffilenewer{figs/Model_fit.svg}{figs/Model_fit.pdf}%
	{inkscape -D figs/Model_fit.svg  -o figs/Model_fit.pdf --export-latex}%
	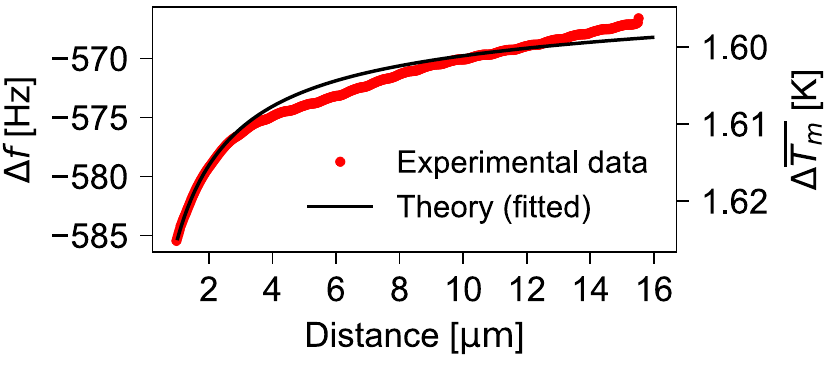
		\vspace{-13.25mm}
		\caption{}
		\label{fig_3:sub-second}
		\vspace{-3.5mm}
	\end{subfigure}
	\caption{ 
		(\subref{fig_3:sub-first}) Raw data scans at room temperature (blue) and under a temperature gradient of $10 \ \mathrm{K}$ (red) for a half sphere displacement velocity of 1 step per second towards the membrane. The $10 \ \mathrm{K}$ gradient scan shows a frequency shift of $\mathrm{\sim\!19 \ Hz}$ confirming near-field thermal coupling, in contrast with the $\Delta T=0 \ \mathrm{K}$ control scan
		(\subref{fig_3:sub-second}) Experimental (red) and theoretical (black) (fitted horizontally and vertically) resonance frequency shift as a function of separation for a $10 \ \mathrm{K}$ thermal gradient. The fit suggests a minimal separation of $\mathrm{\sim\!980 \ nm}$ before sphere-membrane contact. The $-570 \ \mathrm{Hz}$ initial shift at large distances results from far-field radiative coupling.}
	\label{fig_3}
\end{figure}

Raw data scans are presented in Fig.~\ref{fig_3:sub-first}, for a half sphere displacement velocity of 1 step per second towards the membrane. As expected, scans without heating of the half sphere (i.e., $\Delta T = 0 \ \mathrm{K}$) produce no noticeable frequency shift, while scans at $\Delta T = 10 \ \mathrm{K}$, show clear frequency shifts, until contact occurs and resonance mode tracking is lost.  In the case of this data set, a frequency shift of $\mathrm{\sim\!19 \ Hz}$ was measured before contact. 

We compare the experimental frequency shift from Fig.~\ref{fig_3:sub-first} with a theoretical model that combines heat diffusion inside the membrane, multilayer NFHT calculations \cite{St-Gelais_2017,Francoeur_2010}, as well as the effect of non-uniform stress obtained by solving the motion equation \cite{Schmid_2016} in the membrane. As shown in Ref.~\citenum{Zhang_2020}, the radiative thermal coupling properties of square membranes of side length $L$ are well approximated by that of circular membranes of effective radius $r_{eff}=1.252 \frac{L}{2}$. We therefore approximate our square membrane to a circular membrane of radius $r_{eff} = 1.878\ \mathrm{mm}$. The membrane temperature profile $T_m(r)$ is obtained by solving the heat diffusion equation:
 
\begin{equation}
k\nabla^2T_m(r)+\dot{q}_{gen}(r,T_m)=0,
\label{heat_eq}
\end{equation}

\noindent where $k$ is the thermal conductivity of SiN. The internal generation term $\dot{q}_{gen}$ is given as a function of the radius $r$ by NFHT calculations:

\begin{equation}
\begin{split}
\dot{q}_{gen}(r,T_m) & = G_{NF}(r)\cdot\left(T_{sphere}-T_m(r)\right)/t_{SiN}\\ &+\dot{q}_{FF_{front}}(T_m(r))+\dot{q}_{FF_{back}}(T_m(r)),
\end{split}
\label{generation_term}
\end{equation}

\noindent where $T_{sphere}$ is the glass half sphere temperature. In Eq.~\ref{generation_term}, $G_{NF}$, in $\mathrm{W/m^2K}$, accounts for NFHT coupling of evanescent waves and is calculated using the NFHT algorithm given in Ref.~\citenum{St-Gelais_2017} with the permittivity of the glass half sphere taken from Ref.~\citenum{Kitamura_2007} and that of SiN taken from Ref.~\citenum{Cataldo_2012}. Radiative heat transferred in far-field from the back side of the membrane is computed using the emissive power law \cite{Bergman_2011}; $\dot{q}_{FF_{back}}(r)=\varepsilon_{SiN}\sigma(T^4_m(r)-T^4_\infty)/t_{SiN}$, where the emissivity $\varepsilon_{SiN}$ is taken from Ref.~\citenum{Zhang_2020}, $\sigma$ is the Stefan-Boltzmann constant, and $T_\infty=293 \ \mathrm{K}$ is the environment temperature. The far-field radiative heat transferred to the front side of the membrane $\dot{q}_{FF_{front}}$ is calculated using conventional far-field thermal radiation formalism \cite{Bergman_2011} detailed in supplementary section S.1. To account for the sphere curvature, we use the proximity approximation (also known as the Derjaguin approximation \cite{Derjaguin_1956}). Each point on the sphere is approximated to a flat surface with a distance $d(r)=d_0+R-\sqrt{R^2-r^2}$ where $d_0$ is the distance at the sphere apex, and $R$ is the radius of the sphere. The approximation is proven to be valid for NFHT [26], especially for large radius objects like ours. Solving Eq.~\ref{heat_eq}-\ref{generation_term} numerically gives the temperature profile, from which we extract the resonance frequency shift by numerically solving the motion equation in spherical coordinates for a thin membrane under tensile stress \cite{Schmid_2016}:

\begin{equation}
\begin{split}
\frac{1}{r}\frac{\partial}{\partial r}\left(\left(\sigma_0+ \sigma_r(r,T_m)\right)r\frac{\partial u(r,t)}{\partial r}\right)\\
-\rho\frac{\partial^2u(r,t)}{\partial t^2}=0,
\end{split}
\label{motion_eq}
\end{equation}

\noindent where $\sigma_0$ and $\sigma_r(r,T_m)$ are, respectively, the temperature-independent ($\sigma_0\approx100 \ \mathrm{MPa}$) and the position-dependant components of the in-plane stress, $\rho$ is the density of SiN, and $u(r,t)$  represents the out-of-plane displacement of the membrane. The position-dependent component of the in-plane stress ($\sigma_r$) is given as a function of the temperature profile \cite{Sadd_2009}:

\begin{equation}
\begin{split}
\sigma_r(r,T_m)=\alpha E\bigg[\frac{1}{r^2}\int_{0}^{r}r\Delta T_m(r)\mathrm{d} r\\ +\frac{1+\nu}{1-\nu}\frac{\Delta \overline{T_m}}{2}\bigg],
\end{split}
\label{stress_eq}
\end{equation}

\noindent where $\alpha$ is the thermal expansion coefficient of SiN, $E$ is the Young modulus, $\nu$ is the Poisson ratio, and $\Delta T_m=T_m-T_\infty$. $\overline{T_m}$ denotes the temperature average over the membrane area, such that $\Delta\overline{T_m}=\overline{T_m}-T_\infty$. 

In Fig.~\ref{fig_3:sub-second}, the data is in good agreement with the theoretical model (Eq.~\ref{heat_eq}-\ref{stress_eq}). The experimental and theoretical data were fitted together by proceeding to horizontal translation of the experimental data and vertical translation of the theoretical model. The horizontal translation accounts for uncertainty on the initial sphere-membrane distance, while the vertical translation accounts for uncertainty of the glass half sphere surface temperature and for drift in ambient temperature in the vacuum chamber. The fit suggests that we have reached a minimal distance of $\mathrm{\sim\!980 \ nm}$, which is larger than state of the art experiments \cite{Kim_2015,St-Gelais_2016,Salihoglu_2020}. Attraction forces, surface contamination, and sphere-membrane misalignment could be responsible for the imposed gap limit. It is also possible that shocks originating from the inertia drive actuator cause collapsing of the membrane with the half sphere at small separations. Upgrading of our positioners from inertia drive to flexure stage in the near future will eliminate this possibility. At distances larger than $\mathrm{\sim\!3\ \upmu m}$ the frequency shift from the experimental data is slightly larger than the one computed by the theoretical model. A larger infrared material absorption coefficient than considered in the simulation, increasing the contribution of the propagating wave, could explain this discrepancy \cite{St-Gelais_2016}.

\begin{figure}[!htb]
		\centering
		\begin{subfigure}{\columnwidth}
			\def\svgwidth{\columnwidth}
	\executeiffilenewer{figs/allan_dev.svg}{figs/allan_dev.pdf}%
	{inkscape -D figs/allan_dev.svg  -o figs/allan_dev.pdf --export-latex}%
	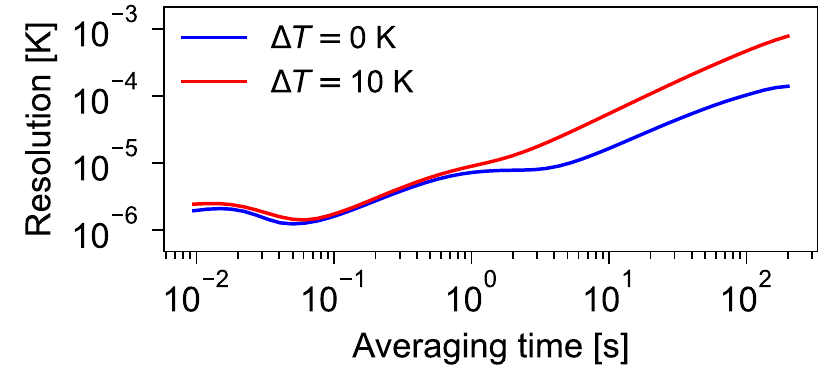

			\vspace{-12.5mm}
			\caption{}
			\label{fig_4:sub-first}
			\vspace{-0.5mm}
		\end{subfigure}
		\begin{subfigure}{\columnwidth}
			\def\svgwidth{\columnwidth}
	\executeiffilenewer{figs/Q_factor.svg}{figs/Q_factor.pdf}%
	{inkscape -D figs/Q_factor.svg  -o figs/Q_factor.pdf --export-latex}%
	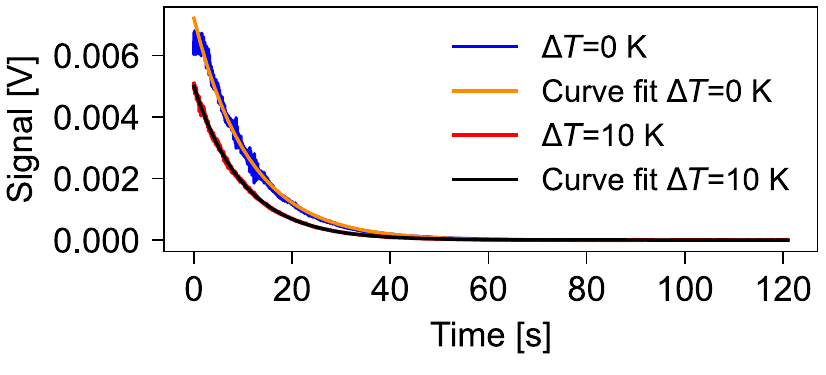

			\vspace{-13.5mm}
			\caption{}
			\label{fig_4:sub-second}
			\vspace{-3.5mm}
		\end{subfigure}
		\caption{ 
			(\subref{fig_4:sub-first}) Resolution of the resonator as a function of averaging time computed from the measured Allan deviation of the PLL frequency signal at both room temperature (blue) and under a temperature gradient of $10 \ \mathrm{K}$ (red). The lowest obtained resolution is $1.2\times10^{-6} \ \mathrm{K}$, which is unprecedented in NFHT measurements. The resolution is limited by drift at higher averaging time, and we note that drift is worse in the presence of a $10 \ \mathrm{K}$ thermal gradient. 
			(\subref{fig_4:sub-second}) Measured mechanical ringdown for mode (2, 2) at room temperature (blue) and under a temperature gradient of 10 K (red). The Q-factor extracted from the exponential fit is comparable at $\Delta T=0 \ \mathrm{K}$ ($2.9\times10^6$) and for $\Delta T=10 \ \mathrm{K}$ ($2.6\times10^6$). }
		\label{fig_4}
	\end{figure}

We find that one of the main advantages of our approach is the very high temperature resolution enabled by the use of a high mechanical quality factor (Q-factor) sensor. By measuring the Allan deviation of the frequency signal and converting it to temperature resolution, we obtain (Fig.~\ref{fig_4:sub-first}) what we believe is an unprecedented resolution in the context of experimental NFHT. We measure a resolution of $1.2\times10^{-6} \ \mathrm{K}$ at low averaging time, while the highest experimentally reported resolution in NFHT measurements is, to the best of our knowledge, $2\times10^{-5} \ \mathrm{K}$ using thermocouples \cite{Kim_2015}. Theoretical analysis claim that bi-material deflection could also achieve similar performances \cite{Lucchesi_2021}, but this was not confirmed experimentally. High resolution in a frequency-shift mechanical sensor directly results from the narrow linewidth of the resonance peaks (i.e., low dissipation, high mechanical Q-factor) \cite{Demir_2021}. Our Q-factor was measured at room temperature and under a temperature gradient of $10 \ \mathrm{K}$ and, in both cases, $\mathrm{Q}>2\times10^6$ was obtained (see Fig.~\ref{fig_4:sub-second}). In contrast with this very high resolution at low averaging time, performances are limited by drift at larger averaging time (see Fig.~\ref{fig_4:sub-first}).  We find that the drift contribution is worse when the half sphere is heated. For this reason, drift currently prevents us from performing conclusive distance scans (i.e., as in Fig.~\ref{fig_3:sub-second}) for $\Delta T>10 \ \mathrm{K}$.

\begin{figure}[!htb]
	\centering
	\begin{subfigure}{\columnwidth}
		\def\svgwidth{\columnwidth}
	\executeiffilenewer{figs/misaligned_lens.svg}{figs/misaligned_lens.pdf}%
	{inkscape -D figs/misaligned_lens.svg  -o figs/misaligned_lens.pdf --export-latex}%
	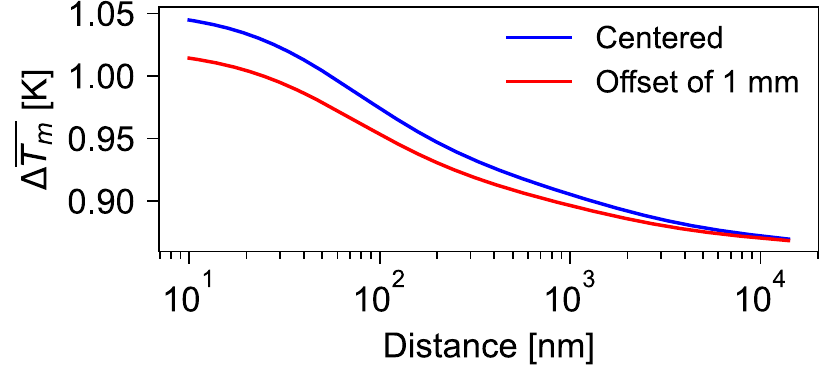
		\vspace{-12.25mm}
		\caption{}
		\label{fig_5:sub-first}
		\vspace{-0.75mm}
	\end{subfigure}
	\begin{subfigure}{\columnwidth}
		\def\svgwidth{\columnwidth}
	\executeiffilenewer{figs/Non-uniform.svg}{figs/Non-uniform.pdf}%
	{inkscape -D figs/Non-uniform.svg  -o figs/Non-uniform.pdf --export-latex}%
	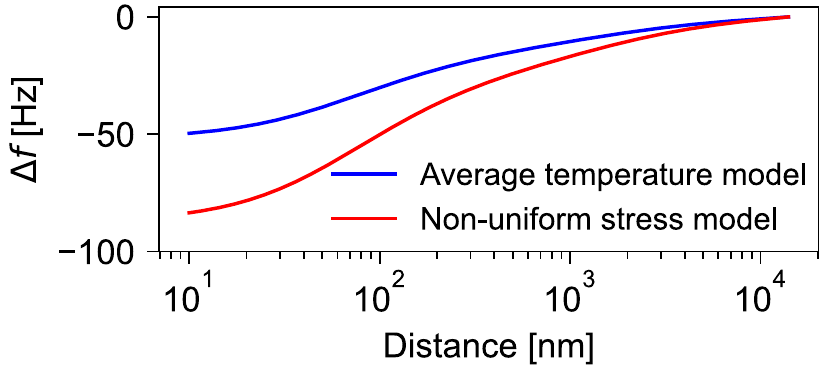
		\vspace{-12.25mm}
		\caption{}
		\label{fig_5:sub-second}
		\vspace{-3.5mm}
	\end{subfigure}
	\caption{Error analysis.
		(\subref{fig_5:sub-first}) Computed near-field coupling of the SiN membrane for a perfectly aligned case (blue), and for a $1 \ \mathrm{mm}$ misalignment between the sphere apex and the membrane center (red). The offset decreases the expected frequency shift, which may affect the minimal distance inferred from Fig.~\ref{fig_3:sub-second}. 
		(\subref{fig_5:sub-second}) Predicted frequency shift from Eq.~\ref{heat_eq}-\ref{stress_eq} (red) compared with the approximation of a uniform membrane temperature (Eq.~\ref{freq_shift}) (blue). The comparison shows a growing discrepancy as we move deeper in the near-field regime, yielding up to a 40\% error at the smallest gap ($10 \ \mathrm{nm}$).}
	\label{fig_5}
\end{figure}

The most important source of uncertainty in our experiment results from the single-axis positionner, which does not allow for active alignment, under vacuum, of the sphere apex with the center of the membrane. In Fig.~\ref{fig_5:sub-first}, we quantify the effect of possible sphere-membrane misalignment by computing the membrane average temperature variation ($\Delta \overline{T_m}$) by solving Eq.~\ref{heat_eq}-\ref{generation_term} with an offset of $1 \ \mathrm{mm}$. We find that this offset decreases the predicted frequency shift, thus potentially explaining the relatively large minimum gap achieved in Fig.~\ref{fig_3:sub-second}. Planned implementation of a 5-axis positioner should alleviate this uncertainty in the near future. 

Rather than solving Eq.~\ref{motion_eq}-\ref{stress_eq} numerically, it would be convenient to express a simple linear relation that converts frequency shift ($\Delta f$) to the average change in membrane temperature ($\Delta \overline{T_m}$). This is done by assuming that $T_m$ is position independent in Eq.~\ref{stress_eq}, from which we obtain the following relation \cite{Blaikie_2019}:

\begin{equation}
\Delta \overline{T_m}=-\frac{2 \sigma_0 (1-\nu)}{\alpha E f_0}\Delta f,
\label{freq_shift}
\end{equation}

\noindent where $f_0$ is the resonance frequency of the membrane at room temperature. In Fig.~\ref{fig_5:sub-second}, we compare the frequency obtained by solving Eq.~\ref{motion_eq} to the one obtained from Eq.~\ref{freq_shift}. By not considering the effect of non-uniform temperature and stress, Eq.~\ref{freq_shift} yields a systematic underestimation of the frequency shift, reaching 40\% at the smallest separation considered here ($10 \ \mathrm{nm}$).     

\begin{figure}[!htb]
	\centering
	\begin{subfigure}{\columnwidth}
		\def\svgwidth{\columnwidth}
	\executeiffilenewer{figs/smaller_membrane.svg}{figs/smaller_membrane.pdf}%
	{inkscape -D figs/smaller_membrane.svg  -o figs/smaller_membrane.pdf --export-latex}%
	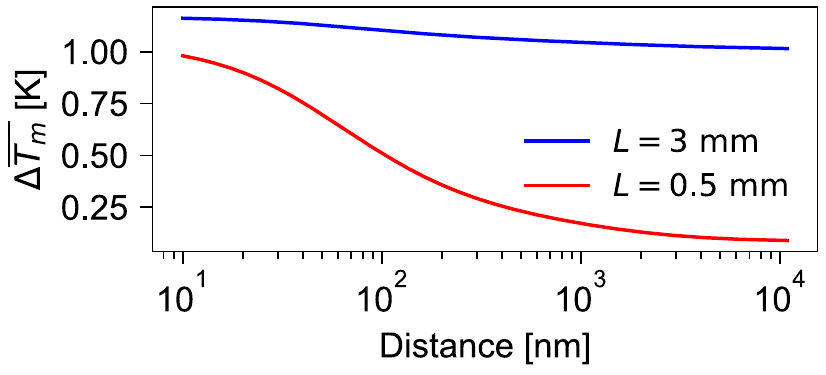

		\vspace{-12.25mm}
		\caption{}
		\label{fig_6:sub-first}
		\vspace{-0.75mm}
	\end{subfigure}
	\begin{subfigure}{\columnwidth}
		\begin{tikzpicture}[scale=1.75]
		\node[font=\fontsize{10pt}{12}\selectfont] at (-0.05,3.3) {$L= 3 \ \mathrm{mm}$};
		\draw[line width=0.08] (-1.2,2.4)--(-0.65,2.4)--(-0.65,2.255)--(-0.85,1.87)--(-1.2,1.87)--(-1.2,2.4);
		\fill[gray, opacity=0.3] (-1.2,2.4)--(-0.65,2.4)--(-0.65,2.255)--(-0.85,1.87)--(-1.2,1.87)--(-1.2,2.4);
		\draw[line width=0.08] (1.1,2.4)--(0.55,2.4)--(0.55,2.255)--(0.75,1.87)--(1.1,1.87)--(1.1,2.4);
		\fill[gray, opacity=0.3] (1.1,2.4)--(0.55,2.4)--(0.55,2.255)--(0.75,1.87)--(1.1,1.87)--(1.1,2.4);
		
		\begin{scope}
		\clip(-1.2,2.55) rectangle (1.1,3.2);

		\draw[line width=0.08] (-5.13,7.65)--(-5.13,7.75);
		\draw[line width=0.08] (5.03,7.65)--(5.03,7.75);
		\draw[line width=0.08] (-5.13,7.75)--(5.03,7.75);
		\draw[line width=0.08,] (5.03,7.65)--(-5.13,7.65) arc(180:360:5.08);
		
		\shade[thin, draw=black, top color=darkred,bottom color=red] 
		(-5.13,7.75)--(-5.13,7.65) arc(180:360:5.08) --(5.03,7.75)-- cycle;
		\fill[white,path fading=south]
		(-1.2,3.1) rectangle (1.1,3.21);
		\fill[white,path fading=west]
		(1,2.55) rectangle (1.11,3.21);
		\fill[white,path fading=east]
		(-1.1,2.55) rectangle (-1.21,3.21);
		\end{scope}
		\node[font=\fontsize{10pt}{12}\selectfont] at (2.5,3.3) {$L= 0.5 \ \mathrm{mm}$};
		
		\draw[line width=0.08] (1.35,2.4)--(1.9,2.4)--(1.9,2.255)--(1.7,1.87)--(1.35,1.87)--(1.35,2.4);
		\fill[gray, opacity=0.3] (1.35,2.4)--(1.9,2.4)--(1.9,2.255)--(1.7,1.87)--(1.35,1.87)--(1.35,2.4);
		\draw[line width=0.08] (3.65,2.4)--(3.1,2.4)--(3.1,2.255)--(3.3,1.87)--(3.65,1.87)--(3.65,2.4);
		\fill[gray, opacity=0.3] (3.65,2.4)--(3.1,2.4)--(3.1,2.255)--(3.3,1.87)--(3.65,1.87)--(3.65,2.4);
		
		
		\begin{scope}
		\clip(1.35,2.55) rectangle (3.65,3.2);

		\draw[line width=0.08] (-27.98,33.05)--(-27.98,33.15);
		\draw[line width=0.08] (32.98,33.05)--(32.98,33.15);
		\draw[line width=0.08] (-27.98,33.15)--(32.98,33.15);
		\draw[line width=0.08,] (32.98,33.05)--(-27.98,33.05) arc(180:360:30.48);
		
		\shade[thin, draw=black, top color=darkred,bottom color=red] 
		(-27.98,33.15)--(-27.98,33.05) arc(180:360:30.48) --(32.98,33.15)-- cycle;
		\fill[white,path fading=south]
		(1.35,3.1) rectangle (3.65,3.21);
		\fill[white,path fading=west]
		(3.55,2.55) rectangle (3.66,3.21);
		\fill[white,path fading=east]
		(1.45,2.55) rectangle (1.34,3.21);
		\end{scope}

		\pgfplotsset{colormap/jet}
		\begin{axis}[ticks=none,
		domain=-2:2,
		view={0}{90},
		width=3.88cm,
		height=1.71cm,
		point meta max= 297.882,
		point meta min=293,
		axis lines=none,
		xshift=-1.2cm,yshift=2.4cm 
		]
		\addplot3[contour filled={number=200}]
		{293};
		\end{axis}
		\pgfplotstableread{Temp_profile_at_10nm_new_emissivity.txt}\datatable
		\pgfplotsset{colormap/jet}
		\begin{axis}
		[ticks=none,
		view={0}{90},
		width=2.78cm,
		height=1.71cm,
		xshift=-0.65cm,yshift=2.4cm,
		point meta max= 297.882,
		point meta min=293,
		axis lines=none,
		]
		\addplot3[mesh/rows=500,mesh/num points=2500,contour filled={number=200}] table [x expr=\thisrowno{0}, y expr=\thisrowno{1}, z expr=\thisrowno{2}] {\datatable};
		\end{axis}
		\begin{axis}[ticks=none,
		view={0}{90},
		width=3.88cm,
		height=1.71cm,
		point meta max= 297.882,
		point meta min=293,
		axis lines=none,
		xshift=1.35cm,yshift=2.4cm 
		]
		\addplot3[contour filled={number=200}]
		{293};
		\end{axis}
		\pgfplotstableread{Temp_profile_at_10nm_new_emissivity_05_mm.txt}\datatable
		\pgfplotsset{colormap/jet}
		\begin{axis}
		[ticks=none,
		view={0}{90},
		width=2.785cm,
		height=1.71cm,
		xshift=1.9cm,yshift=2.4cm,
		point meta max= 297.882,
		point meta min=293,
		axis lines=none,
		]
		\addplot3[mesh/rows=500,mesh/num points=2500,contour filled={number=200}] table [x expr=\thisrowno{0}, y expr=\thisrowno{1}, z expr=\thisrowno{2}] {\datatable};
		\end{axis}
		\end{tikzpicture}
		\vspace{-6mm}
		\caption{}
		\label{fig_6:sub-second}
		\vspace{-0.6mm}
	\end{subfigure}
	\begin{subfigure}{\columnwidth}
		\def\svgwidth{\columnwidth}
	\executeiffilenewer{figs/temp_profile.svg}{figs/temp_profile.pdf}%
	{inkscape -D figs/temp_profile.svg  -o figs/temp_profile.pdf --export-latex}%
	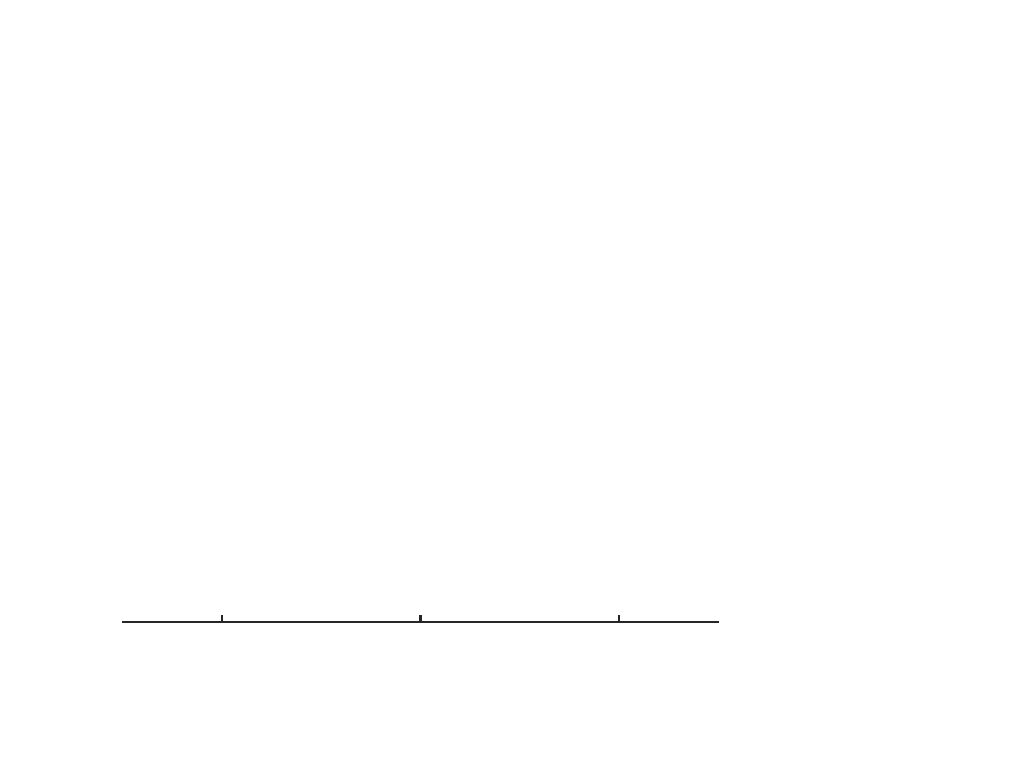

		\vspace{-12.5mm}
		\caption{}
		\label{fig_6:sub-third}
		\vspace{-3.125mm}
	\end{subfigure}
	\caption{
		(\subref{fig_6:sub-first}) Calculated temperature variation of the SiN membrane for the case of a $L= 3 \ \mathrm{mm}$ (blue) and $L= 0.5 \ \mathrm{mm}$ (red) membrane. Larger membranes result in a much smaller temperature increase due to saturation and dilution effects (see main text). 
		(\subref{fig_6:sub-second}) Schematic of the half sphere and membrane showing the computed temperature profile of the SiN resonator at a separation of $10 \ \mathrm{nm}$ in the case of a $L= 3 \ \mathrm{mm}$ (left) and $L= 0.5 \ \mathrm{mm}$ (right) membranes. The curvature of the half sphere is scaled proportionally to the size of the membrane.
		(\subref{fig_6:sub-third}) Two-dimensional view of the simulated SiN membrane temperature profile for a $L= 3 \ \mathrm{mm}$ membrane at a separation of $10 \ \mathrm{nm}$. The temperature at the center of the membrane is close to the sphere temperature, illustrating the saturation effect for large membrane sizes. }
	\label{fig_6}
\end{figure}

Significant optimization of our sphere-membrane approach will be possible in future work. As shown in Fig.~\ref{fig_6:sub-first}, our $L = 3 \ \mathrm{mm}$ membrane results in a very small relative increase in temperature compared to a $L= 0.5 \ \mathrm{mm}$ membrane. A first reason for this effect is weak thermal conduction in large-area SiN membranes. Large-area membranes consequently reach a large temperature in far-field (i.e., a weaker thermal gradient), while small-area membranes remain closer to room temperature by offering a shorter path to the Si frame. Furthermore, at very small distances, near-field thermal radiation is predicted to saturate the membrane temperature at the sphere apex, thus resulting in an even weaker thermal gradient. This effect is worse in large membranes compared to smaller membranes that have higher thermal conduction (see simulated profiles in Figs~\ref{fig_6:sub-second} and \ref{fig_6:sub-third}). Finally, the ratio of area exposed to near-field vs.\ far-field radiation is lower in large membranes, thus greatly diluting the near-field signal. The sole advantage of a large membrane therefore appears to be simplification of the alignment procedure, enabling the use of a single-axis positioner in the present case. 

We have proposed the first experimental demonstration of NFHT measured using nanomechanical resonators, as well as a clear path for substantial improvements by optimizing the resonator dimension. Our approach notably enables the highest reported temperature resolution ($1.2\times10^{-6} \ \mathrm{K}$) in the context of NFHT measurements. We expect that the wide availability of freestanding SiN substrates will make our platform a go-to approach for studying new materials for NFHT. Coupling mechanical oscillators with near-field thermal radiation could also open new interesting research avenues. For example, mechanical oscillations coupled with near-field thermal radiation could be used for implementing systems that rely on temperature oscillations for converting heat to electricity, exploiting for example pyroelectric materials \cite{Fang_2010,Latella_2021}. Ultra-low dissipation of SiN mechanical oscillators could also allow simultaneous characterization of Casimir forces \cite{Obrecht_2007} and NFHT.
\section{Acknowledgement}
This work was funded by the National Science and Engineering Research Council
of Canada (NSERC), the Ontario Graduate Scholarship (OGS), the New Frontiers in Research Fund (NFRF), and by start-up funds from the Faculty of Engineering at the University of Ottawa. 

\begin{suppinfo}
The following files are available free of charge.
\begin{itemize}
	\item Supplementary Information: details on the far-field radiation model.
\end{itemize}

\end{suppinfo}

\bibliography{refs}

\providecommand{\latin}[1]{#1}
\makeatletter
\providecommand{\doi}
  {\begingroup\let\do\@makeother\dospecials
  \catcode`\{=1 \catcode`\}=2 \doi@aux}
\providecommand{\doi@aux}[1]{\endgroup\texttt{#1}}
\makeatother
\providecommand*\mcitethebibliography{\thebibliography}
\csname @ifundefined\endcsname{endmcitethebibliography}
  {\let\endmcitethebibliography\endthebibliography}{}
\begin{mcitethebibliography}{73}
\providecommand*\natexlab[1]{#1}
\providecommand*\mciteSetBstSublistMode[1]{}
\providecommand*\mciteSetBstMaxWidthForm[2]{}
\providecommand*\mciteBstWouldAddEndPuncttrue
  {\def\EndOfBibitem{\unskip.}}
\providecommand*\mciteBstWouldAddEndPunctfalse
  {\let\EndOfBibitem\relax}
\providecommand*\mciteSetBstMidEndSepPunct[3]{}
\providecommand*\mciteSetBstSublistLabelBeginEnd[3]{}
\providecommand*\EndOfBibitem{}
\mciteSetBstSublistMode{f}
\mciteSetBstMaxWidthForm{subitem}{(\alph{mcitesubitemcount})}
\mciteSetBstSublistLabelBeginEnd
  {\mcitemaxwidthsubitemform\space}
  {\relax}
  {\relax}

\bibitem[Laroche \latin{et~al.}(2006)Laroche, Carminati, and
  Greffet]{Laroche_2006}
Laroche,~M.; Carminati,~R.; Greffet,~J.-J. Near-field thermophotovoltaic energy
  conversion. \emph{Journal of Applied Physics} \textbf{2006}, \emph{100},
  063704\relax
\mciteBstWouldAddEndPuncttrue
\mciteSetBstMidEndSepPunct{\mcitedefaultmidpunct}
{\mcitedefaultendpunct}{\mcitedefaultseppunct}\relax
\EndOfBibitem
\bibitem[Milovich \latin{et~al.}(2020)Milovich, Villa, Antolin, Datas, Marti,
  Vaillon, and Francoeur]{Milovich_2020}
Milovich,~D.; Villa,~J.; Antolin,~E.; Datas,~A.; Marti,~A.; Vaillon,~R.;
  Francoeur,~M. {Design of an indium arsenide cell for near-field
  thermophotovoltaic devices}. \emph{Journal of Photonics for Energy}
  \textbf{2020}, \emph{10}, 1 -- 18\relax
\mciteBstWouldAddEndPuncttrue
\mciteSetBstMidEndSepPunct{\mcitedefaultmidpunct}
{\mcitedefaultendpunct}{\mcitedefaultseppunct}\relax
\EndOfBibitem
\bibitem[Fiorino \latin{et~al.}(2018)Fiorino, Zhu, Thompson, Mittapally, Reddy,
  and Meyhofer]{Fiorino_2018}
Fiorino,~A.; Zhu,~L.; Thompson,~D.; Mittapally,~R.; Reddy,~P.; Meyhofer,~E.
  Nanogap near-field thermophotovoltaics. \emph{Nature Nanotechnology}
  \textbf{2018}, \emph{13}, 806–811\relax
\mciteBstWouldAddEndPuncttrue
\mciteSetBstMidEndSepPunct{\mcitedefaultmidpunct}
{\mcitedefaultendpunct}{\mcitedefaultseppunct}\relax
\EndOfBibitem
\bibitem[DiMatteo \latin{et~al.}(2003)DiMatteo, Greiff, Finberg,
  Young‐Waithe, Choy, Masaki, and Fonstad]{DiMatteo_2003}
DiMatteo,~R.~S.; Greiff,~P.; Finberg,~S.~L.; Young‐Waithe,~K.~A.; Choy,~H.
  K.~H.; Masaki,~M.~M.; Fonstad,~C.~G. Micron‐gap ThermoPhotoVoltaics (MTPV).
  \emph{AIP Conference Proceedings} \textbf{2003}, \emph{653}, 232--240\relax
\mciteBstWouldAddEndPuncttrue
\mciteSetBstMidEndSepPunct{\mcitedefaultmidpunct}
{\mcitedefaultendpunct}{\mcitedefaultseppunct}\relax
\EndOfBibitem
\bibitem[Bhatt \latin{et~al.}(2020)Bhatt, Zhao, Roberts, Datta, Mohanty, Lin,
  Hartmann, St-Gelais, Fan, and Lipson]{Bhatt_2020}
Bhatt,~G.~R.; Zhao,~B.; Roberts,~S.; Datta,~I.; Mohanty,~A.; Lin,~T.;
  Hartmann,~J.-M.; St-Gelais,~R.; Fan,~S.; Lipson,~M. Integrated near-field
  thermo-photovoltaics for heat recycling. \emph{Nature Communications}
  \textbf{2020}, \emph{11}, 2545\relax
\mciteBstWouldAddEndPuncttrue
\mciteSetBstMidEndSepPunct{\mcitedefaultmidpunct}
{\mcitedefaultendpunct}{\mcitedefaultseppunct}\relax
\EndOfBibitem
\bibitem[Zhu \latin{et~al.}(2013)Zhu, Otey, and Fan]{Zhu_2013}
Zhu,~L.; Otey,~C.~R.; Fan,~S. Ultrahigh-contrast and large-bandwidth thermal
  rectification in near-field electromagnetic thermal transfer between
  nanoparticles. \emph{Phys. Rev. B} \textbf{2013}, \emph{88}, 184301\relax
\mciteBstWouldAddEndPuncttrue
\mciteSetBstMidEndSepPunct{\mcitedefaultmidpunct}
{\mcitedefaultendpunct}{\mcitedefaultseppunct}\relax
\EndOfBibitem
\bibitem[Biehs \latin{et~al.}(2011)Biehs, Rosa, and Ben-Abdallah]{Biehs_2011}
Biehs,~S.-A.; Rosa,~F. S.~S.; Ben-Abdallah,~P. Modulation of near-field heat
  transfer between two gratings. \emph{Applied Physics Letters} \textbf{2011},
  \emph{98}, 243102\relax
\mciteBstWouldAddEndPuncttrue
\mciteSetBstMidEndSepPunct{\mcitedefaultmidpunct}
{\mcitedefaultendpunct}{\mcitedefaultseppunct}\relax
\EndOfBibitem
\bibitem[Ben-Abdallah \latin{et~al.}(2015)Ben-Abdallah, Belarouci, Frechette,
  and Biehs]{Ben-Abdallah_2015}
Ben-Abdallah,~P.; Belarouci,~A.; Frechette,~L.; Biehs,~S.-A. Heat flux splitter
  for near-field thermal radiation. \emph{Applied Physics Letters}
  \textbf{2015}, \emph{107}, 053109\relax
\mciteBstWouldAddEndPuncttrue
\mciteSetBstMidEndSepPunct{\mcitedefaultmidpunct}
{\mcitedefaultendpunct}{\mcitedefaultseppunct}\relax
\EndOfBibitem
\bibitem[Guha \latin{et~al.}(2012)Guha, Otey, Poitras, Fan, and
  Lipson]{Guha_2012}
Guha,~B.; Otey,~C.; Poitras,~C.~B.; Fan,~S.; Lipson,~M. Near-Field Radiative
  Cooling of Nanostructures. \emph{Nano Letters} \textbf{2012}, \emph{12},
  4546–4550\relax
\mciteBstWouldAddEndPuncttrue
\mciteSetBstMidEndSepPunct{\mcitedefaultmidpunct}
{\mcitedefaultendpunct}{\mcitedefaultseppunct}\relax
\EndOfBibitem
\bibitem[Biehs(2007)]{Biehs_2007}
Biehs,~S.-A. Thermal heat radiation, near-field energy density and near-field
  radiative heat transfer of coated materials. \emph{The European Physical
  Journal B} \textbf{2007}, \emph{58}, 423–431\relax
\mciteBstWouldAddEndPuncttrue
\mciteSetBstMidEndSepPunct{\mcitedefaultmidpunct}
{\mcitedefaultendpunct}{\mcitedefaultseppunct}\relax
\EndOfBibitem
\bibitem[Salihoglu and Xu(2019)Salihoglu, and Xu]{Salihoglu_2019}
Salihoglu,~H.; Xu,~X. Near-field radiative heat transfer enhancement using
  natural hyperbolic material. \emph{Journal of Quantitative Spectroscopy and
  Radiative Transfer} \textbf{2019}, \emph{222-223}, 115--121\relax
\mciteBstWouldAddEndPuncttrue
\mciteSetBstMidEndSepPunct{\mcitedefaultmidpunct}
{\mcitedefaultendpunct}{\mcitedefaultseppunct}\relax
\EndOfBibitem
\bibitem[Basu \latin{et~al.}(2015)Basu, Yang, and Wang]{Basu_2015}
Basu,~S.; Yang,~Y.; Wang,~L. Near-field radiative heat transfer between
  metamaterials coated with silicon carbide thin films. \emph{Applied Physics
  Letters} \textbf{2015}, \emph{106}, 033106\relax
\mciteBstWouldAddEndPuncttrue
\mciteSetBstMidEndSepPunct{\mcitedefaultmidpunct}
{\mcitedefaultendpunct}{\mcitedefaultseppunct}\relax
\EndOfBibitem
\bibitem[Kim \latin{et~al.}(2015)Kim, Song, Fernández-Hurtado, Lee, Jeong,
  Cui, Thompson, Feist, Reid, García-Vidal, and et~al.]{Kim_2015}
Kim,~K.; Song,~B.; Fernández-Hurtado,~V.; Lee,~W.; Jeong,~W.; Cui,~L.;
  Thompson,~D.; Feist,~J.; Reid,~M. T.~H.; García-Vidal,~F.~J.; et~al.,
  Radiative heat transfer in the extreme near field. \emph{Nature}
  \textbf{2015}, \emph{528}, 387–391\relax
\mciteBstWouldAddEndPuncttrue
\mciteSetBstMidEndSepPunct{\mcitedefaultmidpunct}
{\mcitedefaultendpunct}{\mcitedefaultseppunct}\relax
\EndOfBibitem
\bibitem[Francoeur \latin{et~al.}(2008)Francoeur, Mengüç, and
  Vaillon]{Francoeur_2008}
Francoeur,~M.; Mengüç,~M.~P.; Vaillon,~R. Near-field radiative heat transfer
  enhancement via surface phonon polaritons coupling in thin films.
  \emph{Applied Physics Letters} \textbf{2008}, \emph{93}, 043109\relax
\mciteBstWouldAddEndPuncttrue
\mciteSetBstMidEndSepPunct{\mcitedefaultmidpunct}
{\mcitedefaultendpunct}{\mcitedefaultseppunct}\relax
\EndOfBibitem
\bibitem[Jin \latin{et~al.}(2019)Jin, Molesky, Lin, and Rodriguez]{Jin_2019}
Jin,~W.; Molesky,~S.; Lin,~Z.; Rodriguez,~A.~W. Material scaling and
  frequency-selective enhancement of near-field radiative heat transfer for
  lossy metals in two dimensions via inverse design. \emph{Phys. Rev. B}
  \textbf{2019}, \emph{99}, 041403\relax
\mciteBstWouldAddEndPuncttrue
\mciteSetBstMidEndSepPunct{\mcitedefaultmidpunct}
{\mcitedefaultendpunct}{\mcitedefaultseppunct}\relax
\EndOfBibitem
\bibitem[Moncada-Villa \latin{et~al.}(2015)Moncada-Villa, Fern\'andez-Hurtado,
  Garc\'{\i}a-Vidal, Garc\'{\i}a-Mart\'{\i}n, and Cuevas]{Moncada-Villa_2015}
Moncada-Villa,~E.; Fern\'andez-Hurtado,~V.; Garc\'{\i}a-Vidal,~F.~J.;
  Garc\'{\i}a-Mart\'{\i}n,~A.; Cuevas,~J.~C. Magnetic field control of
  near-field radiative heat transfer and the realization of highly tunable
  hyperbolic thermal emitters. \emph{Phys. Rev. B} \textbf{2015}, \emph{92},
  125418\relax
\mciteBstWouldAddEndPuncttrue
\mciteSetBstMidEndSepPunct{\mcitedefaultmidpunct}
{\mcitedefaultendpunct}{\mcitedefaultseppunct}\relax
\EndOfBibitem
\bibitem[Tien \latin{et~al.}(2002)Tien, Lee, and Stretton]{Tien_2002}
Tien,~C.; Lee,~K.; Stretton,~A. Radiation heat transfer. \emph{SFPE Handbook of
  Fire Protection Engineering, Section} \textbf{2002}, \emph{1}\relax
\mciteBstWouldAddEndPuncttrue
\mciteSetBstMidEndSepPunct{\mcitedefaultmidpunct}
{\mcitedefaultendpunct}{\mcitedefaultseppunct}\relax
\EndOfBibitem
\bibitem[Cravalho \latin{et~al.}(1967)Cravalho, Tien, and Caren]{Cravalho_1967}
Cravalho,~E.~G.; Tien,~C.~L.; Caren,~R.~P. Effect of Small Spacings on
  Radiative Transfer Between Two Dielectrics. \emph{Journal of Heat Transfer}
  \textbf{1967}, \emph{89}, 351–358\relax
\mciteBstWouldAddEndPuncttrue
\mciteSetBstMidEndSepPunct{\mcitedefaultmidpunct}
{\mcitedefaultendpunct}{\mcitedefaultseppunct}\relax
\EndOfBibitem
\bibitem[Polder and Van~Hove(1971)Polder, and Van~Hove]{Polder_1971}
Polder,~D.; Van~Hove,~M. Theory of Radiative Heat Transfer between Closely
  Spaced Bodies. \emph{Phys. Rev. B} \textbf{1971}, \emph{4}, 3303--3314\relax
\mciteBstWouldAddEndPuncttrue
\mciteSetBstMidEndSepPunct{\mcitedefaultmidpunct}
{\mcitedefaultendpunct}{\mcitedefaultseppunct}\relax
\EndOfBibitem
\bibitem[Molesky and Jacob(2015)Molesky, and Jacob]{Molesky_2015}
Molesky,~S.; Jacob,~Z. Ideal near-field thermophotovoltaic cells. \emph{Phys.
  Rev. B} \textbf{2015}, \emph{91}, 205435\relax
\mciteBstWouldAddEndPuncttrue
\mciteSetBstMidEndSepPunct{\mcitedefaultmidpunct}
{\mcitedefaultendpunct}{\mcitedefaultseppunct}\relax
\EndOfBibitem
\bibitem[Ilic \latin{et~al.}(2012)Ilic, Jablan, Joannopoulos, Celanovic, and
  Solja\v{c}i\'{c}]{Ilic_2012}
Ilic,~O.; Jablan,~M.; Joannopoulos,~J.~D.; Celanovic,~I.; Solja\v{c}i\'{c},~M.
  Overcoming the black body limit in plasmonic and graphene near-field
  thermophotovoltaic systems. \emph{Opt. Express} \textbf{2012}, \emph{20},
  A366--A384\relax
\mciteBstWouldAddEndPuncttrue
\mciteSetBstMidEndSepPunct{\mcitedefaultmidpunct}
{\mcitedefaultendpunct}{\mcitedefaultseppunct}\relax
\EndOfBibitem
\bibitem[Pendry(1999)]{Pendry_1999}
Pendry,~J.~B. Radiative exchange of heat between nanostructures. \emph{Journal
  of Physics: Condensed Matter} \textbf{1999}, \emph{11}, 6621--6633\relax
\mciteBstWouldAddEndPuncttrue
\mciteSetBstMidEndSepPunct{\mcitedefaultmidpunct}
{\mcitedefaultendpunct}{\mcitedefaultseppunct}\relax
\EndOfBibitem
\bibitem[Carminati and Greffet(1999)Carminati, and Greffet]{Carminati_1999}
Carminati,~R.; Greffet,~J.-J. Near-Field Effects in Spatial Coherence of
  Thermal Sources. \emph{Phys. Rev. Lett.} \textbf{1999}, \emph{82},
  1660--1663\relax
\mciteBstWouldAddEndPuncttrue
\mciteSetBstMidEndSepPunct{\mcitedefaultmidpunct}
{\mcitedefaultendpunct}{\mcitedefaultseppunct}\relax
\EndOfBibitem
\bibitem[Volokitin and Persson(2001)Volokitin, and Persson]{Volokitin_2001}
Volokitin,~A.~I.; Persson,~B. N.~J. Radiative heat transfer between
  nanostructures. \emph{Phys. Rev. B} \textbf{2001}, \emph{63}, 205404\relax
\mciteBstWouldAddEndPuncttrue
\mciteSetBstMidEndSepPunct{\mcitedefaultmidpunct}
{\mcitedefaultendpunct}{\mcitedefaultseppunct}\relax
\EndOfBibitem
\bibitem[Babuty \latin{et~al.}(2013)Babuty, Joulain, Chapuis, Greffet, and
  De~Wilde]{Babuty_2013}
Babuty,~A.; Joulain,~K.; Chapuis,~P.-O.; Greffet,~J.-J.; De~Wilde,~Y. Blackbody
  Spectrum Revisited in the Near Field. \emph{Phys. Rev. Lett.} \textbf{2013},
  \emph{110}, 146103\relax
\mciteBstWouldAddEndPuncttrue
\mciteSetBstMidEndSepPunct{\mcitedefaultmidpunct}
{\mcitedefaultendpunct}{\mcitedefaultseppunct}\relax
\EndOfBibitem
\bibitem[Ben-Abdallah and Biehs(2014)Ben-Abdallah, and
  Biehs]{Ben-Abdallah_2014}
Ben-Abdallah,~P.; Biehs,~S.-A. Near-Field Thermal Transistor. \emph{Phys. Rev.
  Lett.} \textbf{2014}, \emph{112}, 044301\relax
\mciteBstWouldAddEndPuncttrue
\mciteSetBstMidEndSepPunct{\mcitedefaultmidpunct}
{\mcitedefaultendpunct}{\mcitedefaultseppunct}\relax
\EndOfBibitem
\bibitem[Kloppstech \latin{et~al.}(2017)Kloppstech, Könne, Biehs, Rodriguez,
  Worbes, Hellmann, and Kittel]{Kloppstech_2017}
Kloppstech,~K.; Könne,~N.; Biehs,~S.-A.; Rodriguez,~A.~W.; Worbes,~L.;
  Hellmann,~D.; Kittel,~A. Giant heat transfer in the crossover regime between
  conduction and radiation. \emph{Nature Communications} \textbf{2017},
  \emph{8}, 14475\relax
\mciteBstWouldAddEndPuncttrue
\mciteSetBstMidEndSepPunct{\mcitedefaultmidpunct}
{\mcitedefaultendpunct}{\mcitedefaultseppunct}\relax
\EndOfBibitem
\bibitem[Lucchesi \latin{et~al.}(2019)Lucchesi, Cakiroglu, Perez, Taliercio,
  Tournié, Chapuis, and Vaillon]{Lucchesi_2019}
Lucchesi,~C.; Cakiroglu,~D.; Perez,~J.~P.; Taliercio,~T.; Tournié,~E.;
  Chapuis,~P.~O.; Vaillon,~R. Harnessing near-field thermal photons with
  efficient photovoltaic conversion. 2019\relax
\mciteBstWouldAddEndPuncttrue
\mciteSetBstMidEndSepPunct{\mcitedefaultmidpunct}
{\mcitedefaultendpunct}{\mcitedefaultseppunct}\relax
\EndOfBibitem
\bibitem[van Zwol \latin{et~al.}(2012)van Zwol, Ranno, and
  Chevrier]{Van_Zwol_2012}
van Zwol,~P.~J.; Ranno,~L.; Chevrier,~J. Tuning Near Field Radiative Heat Flux
  through Surface Excitations with a Metal Insulator Transition. \emph{Phys.
  Rev. Lett.} \textbf{2012}, \emph{108}, 234301\relax
\mciteBstWouldAddEndPuncttrue
\mciteSetBstMidEndSepPunct{\mcitedefaultmidpunct}
{\mcitedefaultendpunct}{\mcitedefaultseppunct}\relax
\EndOfBibitem
\bibitem[Song \latin{et~al.}(2015)Song, Ganjeh, Sadat, Thompson, Fiorino,
  Fernández-Hurtado, Feist, Garcia-Vidal, Cuevas, Reddy, and
  et~al.]{Song_2015}
Song,~B.; Ganjeh,~Y.; Sadat,~S.; Thompson,~D.; Fiorino,~A.;
  Fernández-Hurtado,~V.; Feist,~J.; Garcia-Vidal,~F.~J.; Cuevas,~J.~C.;
  Reddy,~P.; et~al., Enhancement of near-field radiative heat transfer using
  polar dielectric thin films. \emph{Nature Nanotechnology} \textbf{2015},
  \emph{10}, 253–258\relax
\mciteBstWouldAddEndPuncttrue
\mciteSetBstMidEndSepPunct{\mcitedefaultmidpunct}
{\mcitedefaultendpunct}{\mcitedefaultseppunct}\relax
\EndOfBibitem
\bibitem[Shen \latin{et~al.}(2012)Shen, Mavrokefalos, Sambegoro, and
  Chen]{Shen_2012}
Shen,~S.; Mavrokefalos,~A.; Sambegoro,~P.; Chen,~G. Nanoscale thermal radiation
  between two gold surfaces. \emph{Applied Physics Letters} \textbf{2012},
  \emph{100}, 233114\relax
\mciteBstWouldAddEndPuncttrue
\mciteSetBstMidEndSepPunct{\mcitedefaultmidpunct}
{\mcitedefaultendpunct}{\mcitedefaultseppunct}\relax
\EndOfBibitem
\bibitem[Rousseau \latin{et~al.}(2009)Rousseau, Siria, Jourdan, Volz, Comin,
  Chevrier, and Greffet]{Rousseau_2009}
Rousseau,~E.; Siria,~A.; Jourdan,~G.; Volz,~S.; Comin,~F.; Chevrier,~J.;
  Greffet,~J.-J. Radiative heat transfer at the nanoscale. \emph{Nature
  Photonics} \textbf{2009}, \emph{3}, 514–517\relax
\mciteBstWouldAddEndPuncttrue
\mciteSetBstMidEndSepPunct{\mcitedefaultmidpunct}
{\mcitedefaultendpunct}{\mcitedefaultseppunct}\relax
\EndOfBibitem
\bibitem[Menges \latin{et~al.}(2016)Menges, Dittberner, Novotny, Passarello,
  Parkin, Spieser, Riel, and Gotsmann]{Menges_2016}
Menges,~F.; Dittberner,~M.; Novotny,~L.; Passarello,~D.; Parkin,~S. S.~P.;
  Spieser,~M.; Riel,~H.; Gotsmann,~B. Thermal radiative near field transport
  between vanadium dioxide and silicon oxide across the metal insulator
  transition. \emph{Applied Physics Letters} \textbf{2016}, \emph{108},
  171904\relax
\mciteBstWouldAddEndPuncttrue
\mciteSetBstMidEndSepPunct{\mcitedefaultmidpunct}
{\mcitedefaultendpunct}{\mcitedefaultseppunct}\relax
\EndOfBibitem
\bibitem[Ottens \latin{et~al.}(2011)Ottens, Quetschke, Wise, Alemi, Lundock,
  Mueller, Reitze, Tanner, and Whiting]{Ottens_2011}
Ottens,~R.~S.; Quetschke,~V.; Wise,~S.; Alemi,~A.~A.; Lundock,~R.; Mueller,~G.;
  Reitze,~D.~H.; Tanner,~D.~B.; Whiting,~B.~F. Near-Field Radiative Heat
  Transfer between Macroscopic Planar Surfaces. \emph{Phys. Rev. Lett.}
  \textbf{2011}, \emph{107}, 014301\relax
\mciteBstWouldAddEndPuncttrue
\mciteSetBstMidEndSepPunct{\mcitedefaultmidpunct}
{\mcitedefaultendpunct}{\mcitedefaultseppunct}\relax
\EndOfBibitem
\bibitem[Shi \latin{et~al.}(2019)Shi, Sun, Chen, He, Bao, Evans, and
  He]{Shi_2019}
Shi,~K.; Sun,~Y.; Chen,~Z.; He,~N.; Bao,~F.; Evans,~J.; He,~S. Colossal
  Enhancement of Near-Field Thermal Radiation Across Hundreds of Nanometers
  between Millimeter-Scale Plates through Surface Plasmon and Phonon Polaritons
  Coupling. \emph{Nano Letters} \textbf{2019}, \emph{19}, 8082–8088\relax
\mciteBstWouldAddEndPuncttrue
\mciteSetBstMidEndSepPunct{\mcitedefaultmidpunct}
{\mcitedefaultendpunct}{\mcitedefaultseppunct}\relax
\EndOfBibitem
\bibitem[Lim \latin{et~al.}(2018)Lim, Song, Lee, and Lee]{Lim_2018}
Lim,~M.; Song,~J.; Lee,~S.~S.; Lee,~B.~J. Tailoring near-field thermal
  radiation between metallo-dielectric multilayers using coupled surface
  plasmon polaritons. \emph{Nature Communications} \textbf{2018}, \emph{9},
  4302\relax
\mciteBstWouldAddEndPuncttrue
\mciteSetBstMidEndSepPunct{\mcitedefaultmidpunct}
{\mcitedefaultendpunct}{\mcitedefaultseppunct}\relax
\EndOfBibitem
\bibitem[Kralik \latin{et~al.}(2012)Kralik, Hanzelka, Zobac, Musilova, Fort,
  and Horak]{Kralik_2012}
Kralik,~T.; Hanzelka,~P.; Zobac,~M.; Musilova,~V.; Fort,~T.; Horak,~M. Strong
  Near-Field Enhancement of Radiative Heat Transfer between Metallic Surfaces.
  \emph{Phys. Rev. Lett.} \textbf{2012}, \emph{109}, 224302\relax
\mciteBstWouldAddEndPuncttrue
\mciteSetBstMidEndSepPunct{\mcitedefaultmidpunct}
{\mcitedefaultendpunct}{\mcitedefaultseppunct}\relax
\EndOfBibitem
\bibitem[Ijiro and Yamada(2015)Ijiro, and Yamada]{Ijiro_2015}
Ijiro,~T.; Yamada,~N. Near-field radiative heat transfer between two parallel
  SiO2 plates with and without microcavities. \emph{Applied Physics Letters}
  \textbf{2015}, \emph{106}, 023103\relax
\mciteBstWouldAddEndPuncttrue
\mciteSetBstMidEndSepPunct{\mcitedefaultmidpunct}
{\mcitedefaultendpunct}{\mcitedefaultseppunct}\relax
\EndOfBibitem
\bibitem[Sabbaghi \latin{et~al.}(2020)Sabbaghi, Long, Ying, Lambert, Taylor,
  Messner, and Wang]{Sabbaghi_2020}
Sabbaghi,~P.; Long,~L.; Ying,~X.; Lambert,~L.; Taylor,~S.; Messner,~C.;
  Wang,~L. Super-Planckian radiative heat transfer between macroscale metallic
  surfaces due to near-field and thin-film effects. \emph{Journal of Applied
  Physics} \textbf{2020}, \emph{128}, 025305\relax
\mciteBstWouldAddEndPuncttrue
\mciteSetBstMidEndSepPunct{\mcitedefaultmidpunct}
{\mcitedefaultendpunct}{\mcitedefaultseppunct}\relax
\EndOfBibitem
\bibitem[Tang \latin{et~al.}(2020)Tang, DeSutter, and Francoeur]{Tang_2020}
Tang,~L.; DeSutter,~J.; Francoeur,~M. Near-Field Radiative Heat Transfer
  between Dissimilar Materials Mediated by Coupled Surface Phonon- and
  Plasmon-Polaritons. \emph{ACS Photonics} \textbf{2020}, \emph{7},
  1304–1311\relax
\mciteBstWouldAddEndPuncttrue
\mciteSetBstMidEndSepPunct{\mcitedefaultmidpunct}
{\mcitedefaultendpunct}{\mcitedefaultseppunct}\relax
\EndOfBibitem
\bibitem[DiMatteo \latin{et~al.}(2001)DiMatteo, Greiff, Finberg, Young-Waithe,
  Choy, Masaki, and Fonstad]{DiMatteo_2001}
DiMatteo,~R.~S.; Greiff,~P.; Finberg,~S.~L.; Young-Waithe,~K.~A.; Choy,~H.
  K.~H.; Masaki,~M.~M.; Fonstad,~C.~G. Enhanced photogeneration of carriers in
  a semiconductor via coupling across a nonisothermal nanoscale vacuum gap.
  \emph{Applied Physics Letters} \textbf{2001}, \emph{79}, 1894--1896\relax
\mciteBstWouldAddEndPuncttrue
\mciteSetBstMidEndSepPunct{\mcitedefaultmidpunct}
{\mcitedefaultendpunct}{\mcitedefaultseppunct}\relax
\EndOfBibitem
\bibitem[Hu \latin{et~al.}(2008)Hu, Narayanaswamy, Chen, and Chen]{Hu_2008}
Hu,~L.; Narayanaswamy,~A.; Chen,~X.; Chen,~G. Near-field thermal radiation
  between two closely spaced glass plates exceeding Planck’s blackbody
  radiation law. \emph{Applied Physics Letters} \textbf{2008}, \emph{92},
  133106\relax
\mciteBstWouldAddEndPuncttrue
\mciteSetBstMidEndSepPunct{\mcitedefaultmidpunct}
{\mcitedefaultendpunct}{\mcitedefaultseppunct}\relax
\EndOfBibitem
\bibitem[Ito \latin{et~al.}(2015)Ito, Miura, Iizuka, and Toshiyoshi]{Ito_2015}
Ito,~K.; Miura,~A.; Iizuka,~H.; Toshiyoshi,~H. Parallel-plate submicron gap
  formed by micromachined low-density pillars for near-field radiative heat
  transfer. \emph{Applied Physics Letters} \textbf{2015}, \emph{106},
  083504\relax
\mciteBstWouldAddEndPuncttrue
\mciteSetBstMidEndSepPunct{\mcitedefaultmidpunct}
{\mcitedefaultendpunct}{\mcitedefaultseppunct}\relax
\EndOfBibitem
\bibitem[Zhao \latin{et~al.}(2017)Zhao, Guizal, Zhang, Fan, and
  Antezza]{Zhao_2017}
Zhao,~B.; Guizal,~B.; Zhang,~Z.~M.; Fan,~S.; Antezza,~M. Near-field heat
  transfer between graphene/hBN multilayers. \emph{Phys. Rev. B} \textbf{2017},
  \emph{95}, 245437\relax
\mciteBstWouldAddEndPuncttrue
\mciteSetBstMidEndSepPunct{\mcitedefaultmidpunct}
{\mcitedefaultendpunct}{\mcitedefaultseppunct}\relax
\EndOfBibitem
\bibitem[Lucchesi \latin{et~al.}(2021)Lucchesi, Vaillon, and
  Chapuis]{Lucchesi_2021}
Lucchesi,~C.; Vaillon,~R.; Chapuis,~P.-O. Radiative heat transfer at the
  nanoscale: experimental trends and challenges. \emph{Nanoscale Horizons}
  \textbf{2021}, \emph{6}, 201–208\relax
\mciteBstWouldAddEndPuncttrue
\mciteSetBstMidEndSepPunct{\mcitedefaultmidpunct}
{\mcitedefaultendpunct}{\mcitedefaultseppunct}\relax
\EndOfBibitem
\bibitem[Nika and Balandin(2012)Nika, and Balandin]{Nika_2012}
Nika,~D.~L.; Balandin,~A.~A. Two-dimensional phonon transport in graphene.
  \emph{Journal of Physics: Condensed Matter} \textbf{2012}, \emph{24},
  233203\relax
\mciteBstWouldAddEndPuncttrue
\mciteSetBstMidEndSepPunct{\mcitedefaultmidpunct}
{\mcitedefaultendpunct}{\mcitedefaultseppunct}\relax
\EndOfBibitem
\bibitem[Shi \latin{et~al.}(2017)Shi, Bao, and He]{Shi_2017}
Shi,~K.; Bao,~F.; He,~S. Enhanced Near-Field Thermal Radiation Based on
  Multilayer Graphene-hBN Heterostructures. \emph{ACS Photonics} \textbf{2017},
  \emph{4}, 971–978\relax
\mciteBstWouldAddEndPuncttrue
\mciteSetBstMidEndSepPunct{\mcitedefaultmidpunct}
{\mcitedefaultendpunct}{\mcitedefaultseppunct}\relax
\EndOfBibitem
\bibitem[Creemer \latin{et~al.}(2010)Creemer, Helveg, Kooyman, Molenbroek,
  Zandbergen, and Sarro]{Creemer_2010}
Creemer,~J.~F.; Helveg,~S.; Kooyman,~P.~J.; Molenbroek,~A.~M.;
  Zandbergen,~H.~W.; Sarro,~P.~M. A MEMS Reactor for Atomic-Scale Microscopy of
  Nanomaterials Under Industrially Relevant Conditions. \emph{Journal of
  Microelectromechanical Systems} \textbf{2010}, \emph{19}, 254–264\relax
\mciteBstWouldAddEndPuncttrue
\mciteSetBstMidEndSepPunct{\mcitedefaultmidpunct}
{\mcitedefaultendpunct}{\mcitedefaultseppunct}\relax
\EndOfBibitem
\bibitem[Thompson \latin{et~al.}(2008)Thompson, Zwickl, Jayich, Marquardt,
  Girvin, and Harris]{Thompson_2008}
Thompson,~J.~D.; Zwickl,~B.~M.; Jayich,~A.~M.; Marquardt,~F.; Girvin,~S.~M.;
  Harris,~J. G.~E. Strong dispersive coupling of a high-finesse cavity to a
  micromechanical membrane. \emph{Nature} \textbf{2008}, \emph{452},
  72–75\relax
\mciteBstWouldAddEndPuncttrue
\mciteSetBstMidEndSepPunct{\mcitedefaultmidpunct}
{\mcitedefaultendpunct}{\mcitedefaultseppunct}\relax
\EndOfBibitem
\bibitem[Wilson \latin{et~al.}(2009)Wilson, Regal, Papp, and
  Kimble]{Wilson_2009}
Wilson,~D.~J.; Regal,~C.~A.; Papp,~S.~B.; Kimble,~H.~J. Cavity Optomechanics
  with Stoichiometric SiN Films. \emph{Phys. Rev. Lett.} \textbf{2009},
  \emph{103}, 207204\relax
\mciteBstWouldAddEndPuncttrue
\mciteSetBstMidEndSepPunct{\mcitedefaultmidpunct}
{\mcitedefaultendpunct}{\mcitedefaultseppunct}\relax
\EndOfBibitem
\bibitem[Norte \latin{et~al.}(2016)Norte, Moura, and Gr\"oblacher]{Norte_2016}
Norte,~R.~A.; Moura,~J.~P.; Gr\"oblacher,~S. Mechanical Resonators for Quantum
  Optomechanics Experiments at Room Temperature. \emph{Phys. Rev. Lett.}
  \textbf{2016}, \emph{116}, 147202\relax
\mciteBstWouldAddEndPuncttrue
\mciteSetBstMidEndSepPunct{\mcitedefaultmidpunct}
{\mcitedefaultendpunct}{\mcitedefaultseppunct}\relax
\EndOfBibitem
\bibitem[Rugar \latin{et~al.}(1989)Rugar, Mamin, and Guethner]{Rugar_1989}
Rugar,~D.; Mamin,~H.~J.; Guethner,~P. Improved fiber‐optic interferometer for
  atomic force microscopy. \emph{Applied Physics Letters} \textbf{1989},
  \emph{55}, 2588--2590\relax
\mciteBstWouldAddEndPuncttrue
\mciteSetBstMidEndSepPunct{\mcitedefaultmidpunct}
{\mcitedefaultendpunct}{\mcitedefaultseppunct}\relax
\EndOfBibitem
\bibitem[{Zhang} \latin{et~al.}(2019){Zhang}, {Giroux}, {Nour}, and
  {St-Gelais}]{Zhang_2019}
{Zhang},~C.; {Giroux},~M.; {Nour},~T.~A.; {St-Gelais},~R. Thermal radiation
  sensing using high mechanical Q-factor silicon nitride membranes. 2019 IEEE
  SENSORS. 2019; pp 1--4\relax
\mciteBstWouldAddEndPuncttrue
\mciteSetBstMidEndSepPunct{\mcitedefaultmidpunct}
{\mcitedefaultendpunct}{\mcitedefaultseppunct}\relax
\EndOfBibitem
\bibitem[Sadeghi \latin{et~al.}(2020)Sadeghi, Demir, Villanueva, K\"ahler, and
  Schmid]{Sadeghi_2020}
Sadeghi,~P.; Demir,~A.; Villanueva,~L.~G.; K\"ahler,~H.; Schmid,~S. Frequency
  fluctuations in nanomechanical silicon nitride string resonators. \emph{Phys.
  Rev. B} \textbf{2020}, \emph{102}, 214106\relax
\mciteBstWouldAddEndPuncttrue
\mciteSetBstMidEndSepPunct{\mcitedefaultmidpunct}
{\mcitedefaultendpunct}{\mcitedefaultseppunct}\relax
\EndOfBibitem
\bibitem[Lipoma(1971)]{Lipoma_1971}
Lipoma,~P. Wein bridge oscillator circuit. \textbf{1971}, \relax
\mciteBstWouldAddEndPunctfalse
\mciteSetBstMidEndSepPunct{\mcitedefaultmidpunct}
{}{\mcitedefaultseppunct}\relax
\EndOfBibitem
\bibitem[Zwickl \latin{et~al.}(2008)Zwickl, Shanks, Jayich, Yang,
  Bleszynski~Jayich, Thompson, and Harris]{Zwickl_2008}
Zwickl,~B.~M.; Shanks,~W.~E.; Jayich,~A.~M.; Yang,~C.;
  Bleszynski~Jayich,~A.~C.; Thompson,~J.~D.; Harris,~J. G.~E. High quality
  mechanical and optical properties of commercial silicon nitride membranes.
  \emph{Applied Physics Letters} \textbf{2008}, \emph{92}, 103125\relax
\mciteBstWouldAddEndPuncttrue
\mciteSetBstMidEndSepPunct{\mcitedefaultmidpunct}
{\mcitedefaultendpunct}{\mcitedefaultseppunct}\relax
\EndOfBibitem
\bibitem[St-Gelais \latin{et~al.}(2017)St-Gelais, Bhatt, Zhu, Fan, and
  Lipson]{St-Gelais_2017}
St-Gelais,~R.; Bhatt,~G.~R.; Zhu,~L.; Fan,~S.; Lipson,~M. Hot Carrier-Based
  Near-Field Thermophotovoltaic Energy Conversion. \emph{ACS Nano}
  \textbf{2017}, \emph{11}, 3001–3009\relax
\mciteBstWouldAddEndPuncttrue
\mciteSetBstMidEndSepPunct{\mcitedefaultmidpunct}
{\mcitedefaultendpunct}{\mcitedefaultseppunct}\relax
\EndOfBibitem
\bibitem[Francoeur \latin{et~al.}(2010)Francoeur, Mengü{\c{c}}, and
  Vaillon]{Francoeur_2010}
Francoeur,~M.; Mengü{\c{c}},~M.~P.; Vaillon,~R. Spectral tuning of near-field
  radiative heat flux between two thin silicon carbide films. \emph{Journal of
  Physics D: Applied Physics} \textbf{2010}, \emph{43}, 075501\relax
\mciteBstWouldAddEndPuncttrue
\mciteSetBstMidEndSepPunct{\mcitedefaultmidpunct}
{\mcitedefaultendpunct}{\mcitedefaultseppunct}\relax
\EndOfBibitem
\bibitem[Schmid \latin{et~al.}(2016)Schmid, Villanueva, and
  Roukes]{Schmid_2016}
Schmid,~S.; Villanueva,~L.~G.; Roukes,~M.~L. \emph{Fundamentals of
  nanomechanical resonators}; Springer, 2016; Vol.~49\relax
\mciteBstWouldAddEndPuncttrue
\mciteSetBstMidEndSepPunct{\mcitedefaultmidpunct}
{\mcitedefaultendpunct}{\mcitedefaultseppunct}\relax
\EndOfBibitem
\bibitem[Zhang \latin{et~al.}(2020)Zhang, Giroux, Nour, and
  St-Gelais]{Zhang_2020}
Zhang,~C.; Giroux,~M.; Nour,~T.~A.; St-Gelais,~R. Radiative Heat Transfer in
  Freestanding Silicon Nitride Membranes. \emph{Phys. Rev. Applied}
  \textbf{2020}, \emph{14}, 024072\relax
\mciteBstWouldAddEndPuncttrue
\mciteSetBstMidEndSepPunct{\mcitedefaultmidpunct}
{\mcitedefaultendpunct}{\mcitedefaultseppunct}\relax
\EndOfBibitem
\bibitem[Kitamura \latin{et~al.}(2007)Kitamura, Pilon, and
  Jonasz]{Kitamura_2007}
Kitamura,~R.; Pilon,~L.; Jonasz,~M. Optical constants of silica glass from
  extreme ultraviolet to far infrared at near room temperature. \emph{Appl.
  Opt.} \textbf{2007}, \emph{46}, 8118--8133\relax
\mciteBstWouldAddEndPuncttrue
\mciteSetBstMidEndSepPunct{\mcitedefaultmidpunct}
{\mcitedefaultendpunct}{\mcitedefaultseppunct}\relax
\EndOfBibitem
\bibitem[Cataldo \latin{et~al.}(2012)Cataldo, Beall, Cho, McAndrew, Niemack,
  and Wollack]{Cataldo_2012}
Cataldo,~G.; Beall,~J.~A.; Cho,~H.-M.; McAndrew,~B.; Niemack,~M.~D.;
  Wollack,~E.~J. Infrared dielectric properties of low-stress silicon nitride.
  \emph{Opt. Lett.} \textbf{2012}, \emph{37}, 4200--4202\relax
\mciteBstWouldAddEndPuncttrue
\mciteSetBstMidEndSepPunct{\mcitedefaultmidpunct}
{\mcitedefaultendpunct}{\mcitedefaultseppunct}\relax
\EndOfBibitem
\bibitem[Bergman \latin{et~al.}(2011)Bergman, Incropera, DeWitt, and
  Lavine]{Bergman_2011}
Bergman,~T.~L.; Incropera,~F.~P.; DeWitt,~D.~P.; Lavine,~A.~S.
  \emph{Fundamentals of heat and mass transfer}; John Wiley \& Sons, 2011\relax
\mciteBstWouldAddEndPuncttrue
\mciteSetBstMidEndSepPunct{\mcitedefaultmidpunct}
{\mcitedefaultendpunct}{\mcitedefaultseppunct}\relax
\EndOfBibitem
\bibitem[Derjaguin \latin{et~al.}(1956)Derjaguin, Abrikosova, and
  Lifshitz]{Derjaguin_1956}
Derjaguin,~B.~V.; Abrikosova,~I.~I.; Lifshitz,~E.~M. Direct measurement of
  molecular attraction between solids separated by a narrow gap. \emph{Q. Rev.
  Chem. Soc.} \textbf{1956}, \emph{10}, 295--329\relax
\mciteBstWouldAddEndPuncttrue
\mciteSetBstMidEndSepPunct{\mcitedefaultmidpunct}
{\mcitedefaultendpunct}{\mcitedefaultseppunct}\relax
\EndOfBibitem
\bibitem[Sadd(2009)]{Sadd_2009}
Sadd,~M.~H. \emph{Elasticity: theory, applications, and numerics}; Academic
  Press, 2009\relax
\mciteBstWouldAddEndPuncttrue
\mciteSetBstMidEndSepPunct{\mcitedefaultmidpunct}
{\mcitedefaultendpunct}{\mcitedefaultseppunct}\relax
\EndOfBibitem
\bibitem[St-Gelais \latin{et~al.}(2016)St-Gelais, Zhu, Fan, and
  Lipson]{St-Gelais_2016}
St-Gelais,~R.; Zhu,~L.; Fan,~S.; Lipson,~M. Near-field radiative heat transfer
  between parallel structures in the deep subwavelength regime. \emph{Nature
  Nanotechnology} \textbf{2016}, \emph{11}, 515–519\relax
\mciteBstWouldAddEndPuncttrue
\mciteSetBstMidEndSepPunct{\mcitedefaultmidpunct}
{\mcitedefaultendpunct}{\mcitedefaultseppunct}\relax
\EndOfBibitem
\bibitem[Salihoglu \latin{et~al.}(2020)Salihoglu, Nam, Traverso, Segovia,
  Venuthurumilli, Liu, Wei, Li, and Xu]{Salihoglu_2020}
Salihoglu,~H.; Nam,~W.; Traverso,~L.; Segovia,~M.; Venuthurumilli,~P.~K.;
  Liu,~W.; Wei,~Y.; Li,~W.; Xu,~X. Near-Field Thermal Radiation between Two
  Plates with Sub-10 nm Vacuum Separation. \emph{Nano Letters} \textbf{2020},
  \emph{20}, 6091–6096\relax
\mciteBstWouldAddEndPuncttrue
\mciteSetBstMidEndSepPunct{\mcitedefaultmidpunct}
{\mcitedefaultendpunct}{\mcitedefaultseppunct}\relax
\EndOfBibitem
\bibitem[Demir(2021)]{Demir_2021}
Demir,~A. Understanding fundamental trade-offs in nanomechanical resonant
  sensors. \emph{Journal of Applied Physics} \textbf{2021}, \emph{129},
  044503\relax
\mciteBstWouldAddEndPuncttrue
\mciteSetBstMidEndSepPunct{\mcitedefaultmidpunct}
{\mcitedefaultendpunct}{\mcitedefaultseppunct}\relax
\EndOfBibitem
\bibitem[Blaikie \latin{et~al.}(2019)Blaikie, Miller, and
  Alemán]{Blaikie_2019}
Blaikie,~A.; Miller,~D.; Alemán,~B.~J. A fast and sensitive room-temperature
  graphene nanomechanical bolometer. \emph{Nature Communications}
  \textbf{2019}, \emph{10}, 4726\relax
\mciteBstWouldAddEndPuncttrue
\mciteSetBstMidEndSepPunct{\mcitedefaultmidpunct}
{\mcitedefaultendpunct}{\mcitedefaultseppunct}\relax
\EndOfBibitem
\bibitem[Fang \latin{et~al.}(2010)Fang, Frederich, and Pilon]{Fang_2010}
Fang,~J.; Frederich,~H.; Pilon,~L. Harvesting Nanoscale Thermal Radiation Using
  Pyroelectric Materials. \emph{Journal of Heat Transfer} \textbf{2010},
  \emph{132}, 092701\relax
\mciteBstWouldAddEndPuncttrue
\mciteSetBstMidEndSepPunct{\mcitedefaultmidpunct}
{\mcitedefaultendpunct}{\mcitedefaultseppunct}\relax
\EndOfBibitem
\bibitem[Latella and Ben-Abdallah(2021)Latella, and Ben-Abdallah]{Latella_2021}
Latella,~I.; Ben-Abdallah,~P. Graphene-based autonomous pyroelectric system for
  near-field energy conversion. 2021\relax
\mciteBstWouldAddEndPuncttrue
\mciteSetBstMidEndSepPunct{\mcitedefaultmidpunct}
{\mcitedefaultendpunct}{\mcitedefaultseppunct}\relax
\EndOfBibitem
\bibitem[Obrecht \latin{et~al.}(2007)Obrecht, Wild, Antezza, Pitaevskii,
  Stringari, and Cornell]{Obrecht_2007}
Obrecht,~J.~M.; Wild,~R.~J.; Antezza,~M.; Pitaevskii,~L.~P.; Stringari,~S.;
  Cornell,~E.~A. Measurement of the Temperature Dependence of the
  Casimir-Polder Force. \emph{Phys. Rev. Lett.} \textbf{2007}, \emph{98},
  063201\relax
\mciteBstWouldAddEndPuncttrue
\mciteSetBstMidEndSepPunct{\mcitedefaultmidpunct}
{\mcitedefaultendpunct}{\mcitedefaultseppunct}\relax
\EndOfBibitem
\end{mcitethebibliography}

\renewcommand{\citenumfont}[1]{S#1} 
\renewcommand{\bibnumfmt}[1]{(S#1)} 
\renewcommand\thefigure{S\arabic{figure}}
\renewcommand\theequation{S\arabic{equation}}
\setcounter{figure}{0} 
\setcounter{equation}{0} 

\onecolumn
\begin{center}
	\section{(Supplementary Information)}
\end{center}
\subsection{S.1 Far-field radiation model}
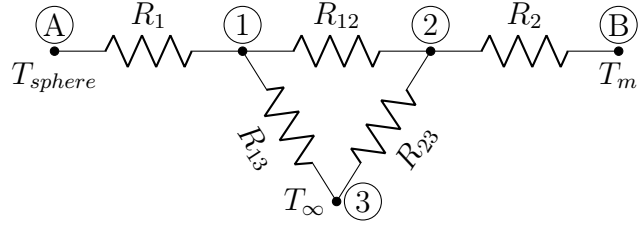
\begin{figure}[H]
	\centering
	\begin{circuitikz}[american]
		\ctikzset{tripoles/mos style/arrows}
		\draw (0,0.35) node[]{A}
		(0,0.35) circle (7pt)
		(0,0) node[circ]{} node[below]{$T_{sphere}$}
		(0,0) to [R=$R_1$] (2.5,0) 
		(2.5,0) node[circ]{} 
		(2.5,0.35) node[]{1}
		(2.5,0.35) circle (7pt)
		(2.5,0)to [R=$R_{12}$] (5,0) 
		(5,0.35) node[]{2}
		(5,0) node[circ]{}
		(5,0.35) circle (7pt)
		(5,0)to [R=$R_{2}$] (7.5,0) 
		(7.5,0.35) node[]{B}
		(7.5,0.35) circle (7pt)
		(7.5,0) node[circ]{} node[below]{$T_m$}
		(3.75,-2)to [R=$R_{13}$] (2.5,0)
		(4.1,-2) node[]{3}
		(4.1,-2) circle (7pt)
		(3.75,-2) node[circ]{} node[left]{$T_{\infty}$}
		(5,0)to [R=$R_{23}$] (3.75,-2)
		
		;
	\end{circuitikz}
	
	\caption{Equivalent thermal circuit of the far-field radiation between the half sphere (point A), the membrane (point B) and the environment (point 3). In this case, the environment is considered a blackbody. $R_1$ and $R_2$ are the surface resistances and $R_{12}$, $R_{13}$ and $R_{23}$ are the space resistances accounting for the view factors.}
	\label{fig_S1}
\end{figure}

The far-field radiative heat transferred to the front side of the membrane $\dot{q}_{FF_{front}}$ can be computed from the equivalent thermal circuit, represented in Fig.~\ref{fig_S1}, for the three surrounding surfaces: the environment (point 3), the half sphere (point A) and the SiN membrane (point B). This model approximates the environment to a blackbody, consequently, neglecting the photons reflected on the environment surface travelling back to the half sphere or membrane surface.

$R_1$ and $R_2$ are, respectively, the surface resistances of the half sphere and of the SiN membrane, calculated as:$^{\mathrm{S1}}$

\begin{equation}
R_1=\frac{1-\varepsilon_{BK7}}{A_1\varepsilon_{BK7}},
\label{R_1_eq}
\end{equation}

\begin{equation}
R_2=\frac{1-\varepsilon_{SiN}}{A_2\varepsilon_{SiN}},
\label{R_2_eq}
\end{equation}

\noindent where $\varepsilon_{BK7}$ is the emissivity of the half sphere, $A_1$ is the surface area of the sphere (the area of a full sphere is used to agree with the view factor model, see supplementary subsection S.1.1), and $A_2$ is the surface area of the SiN membrane. $R_{12}$, $R_{13}$ and $R_{23}$ are the space resistances between the half sphere, the SiN membrane, and the environment accounting for the view factors (supplementary subsection S.1.1). The space resistances are given as:

\begin{equation}
R_{12}=\frac{1}{A_1F_{12}},
\label{R_12_eq}
\end{equation}

\begin{equation}
R_{13}=\frac{1}{A_1F_{13}},
\label{R_13_eq}
\end{equation}

\begin{equation}
R_{23}=\frac{1}{A_2F_{23}},
\label{R_23_eq}
\end{equation}

\noindent where $F_{12}$ is the view factor from the half sphere to the SiN membrane, $F_{13}$ is the view factor from the half sphere to the environment, and $F_{23}$ is the view factor from the SiN membrane to the environment. At nodes A, B, and 3 the equivalent potential applied to the circuit is equal to the blackbody emissive powers ($E_b$):

\begin{equation}
E_{bA}=\sigma T^4_{sphere},
\label{E_bA_eq}
\end{equation}

\begin{equation}
E_{bB}=\sigma T^4_{m},
\label{E_bB_eq}
\end{equation}

\begin{equation}
E_{b3}=\sigma T^4_\infty.
\label{E_b3_eq}
\end{equation}

A relation for the potential at node 2 ($P_2$) can be derived from the application of the Kirchhoff current law at node 1 and 2, yielding the following equations:

\begin{equation}
\frac{E_{bA}-P_1}{R_1}+\frac{P_2-P_1}{R_{12}}+\frac{E_{b3}-P_1}{R_{13}}=0,
\label{Kirchhoff_node_1}
\end{equation}

\begin{equation}
\frac{E_{bB}-P_2}{R_2}+\frac{P_1-P_2}{R_{12}}+\frac{E_{b3}-P_2}{R_{23}}=0,
\label{Kirchhoff_node_2}
\end{equation}

\noindent where $P_1$ is the potential at node 1. Solving this analytically gives us the following relation for the potential at node 2:

\begin{equation}
P_2=\frac{
	\splitfrac{
		E_{bA}R_{13}R_2R_{23}+E_{bB}R_{1}R_{12}R_{23}+E_{bB}R_{1}R_{13}R_{23}+E_{bB}R_{12}R_{13}R_{23}
	}{
		+E_{b3}R_{1}R_{12}R_{2}+E_{b3}R_{1}R_{13}R_{2}+E_{b3}R_{1}R_{2}R_{23}+E_{b3}R_{12}R_{13}R_{2}
	}
}{
	\splitfrac{
		R_1R_{12}R_{2}+R_1R_{12}R_{23}+R_1R_{13}R_{2}+R_1R_{13}R_{23}
	}{
		+R_1R_{2}R_{23}+R_2R_{12}R_{13}+R_{12}R_{13}R_{23}+R_{13}R_{2}R_{23}
	}
}.
\label{P_2_eq}
\end{equation}

Using Ohm’s law at the resistance $R_2$ and the formula derived for the potential at node 2, one can compute the heat absorbed by unit volume due to radiation at the front side of the SiN membrane $\dot{q}_{FF_{front}}$:

\begin{equation}
\dot{q}_{FF_{front}}=\frac{P_2-E_{bB}}{R_2t_{SiN}}.
\label{q_ff_front_eq}
\end{equation}

\subsection{S.1.1 View factor model}                                 

The view factor from the half sphere to the SiN membrane is computed using the tabulated relation from Ref.~S2 for a sphere of radius $R$ and a disk of radius $r_{eff}$ separated by a distance $d_0$, yielding the following equation: 

\begin{equation}
F_{12}=\frac{1}{2}\left(1-\frac{1}{\sqrt{1+\frac{1}{h^2}}}\right),
\label{F_12_eq}
\end{equation}

\noindent where $h=\frac{d_0+R}{r_{eff}}$. Using the principle of reciprocity,

\begin{equation}
A_1F_{12}=A_2F_{21},
\label{reciprocity}
\end{equation}

\noindent allows us to find the view factor for the radiation from the membrane to the half sphere $F_{21}$. Finally, the view factors for the radiation going from the sphere to the environment ($F_{13}$) and from the membrane to the environment ($F_{23}$) are computed using the enclosure summation:

\begin{equation}
F_{13}=1-F_{12},
\label{F_13_eq}
\end{equation}

\begin{equation}
F_{23}=1-F_{21}.
\label{F_23_eq}
\end{equation}

\section{References}
\begin{enumerate}[label=(S\arabic*)]
	\item Bergman, T. L.; Incropera, F. P.; DeWitt, D. P.; Lavine, A. S. \textit{Fundamentals of heat and mass transfer}; John Wiley \& Sons, 2011.
	\item Howell, J. R.; Mengüç, M. P. Radiative transfer configuration factor catalog: A listing of relations for common geometries. \textit{Journal of Quantitative Spectroscopy and Radiative Transfer} \textbf{2011}, \textit{112}, 910–912.
\end{enumerate}

\end{document}